# The mechanical and electrochemical properties of polyaniline-coated carbon nanotube mat

Wei Tan[1,2], Joe C. Stallard[1], Changshin Jo[1,3], Michael F. L. De Volder[1], Norman A. Fleck[1]*

[1] Engineering Department, University of Cambridge, Trumpington Street, Cambridge, CB2 1PZ, UK.
[2] School of Engineering and Materials Science, Queen Mary University London, London, E1 4NS, UK.
[3] School of Chemical Engineering & Materials Science, Chung-Ang University, 84 Heukseok-ro, Dongjak-gu, Seoul, 06974, Republic of Korea

## Abstract

The measured capacitance, modulus and strength of carbon nanotube-polyaniline (CNT-PANI) composite electrodes render them promising candidates for structural energy storage devices. Here, CNT-PANI composite electrodes are manufactured with electrodeposition of PANI onto the bundle network of CNT mats produced via a floating catalyst chemical vapour deposition process. PANI comprises 0% to 30% by volume of the electrode. The composition, modulus, strength and capacitance of the electrodes is measured in the initial state, after the first charge, and after 1000 charge/discharge cycles. Electrode modulus and strength increase with increasing CNT volume fraction; in contrast, the capacitance increases with increasing PANI mass. Charging or cycling reduce the electrode modulus and strength due to a decrease in CNT bundle volume fraction caused by swelling; the electrode capacitance also decreases due to a reduction in PANI mass. A micromechanical model is able to predict the stress-strain response of pre-charged and cycled electrodes, based upon their measured composition after pre-charging and cycling. The electrodes possess up to 63% of their theoretical capacitance, and their tensile strengths are comparable to those of engineering alloys. Their capacitance and strength decrease by less than 15% after the application of 1000 charge/discharge cycles. These properties illustrate their potential as structural energy storage devices.

Keywords: Pseudocapacitor; Carbon nanotubes; Polyaniline; Mechanical properties; Electrochemical properties.

---

*Corresponding author. Tel: +44 (0) 1223 748240. Email: naf1@eng.cam.ac.uk (Norman Fleck)

# 1. Introduction

Supercapacitors combine high specific power with cyclic stability. Consequently, they find application in regenerative braking systems and in other applications that require high power, short-term energy storage [1,2]. The configuration of a typical supercapacitor is drawn in Figure 1(a). Two current collectors are coated with electrode materials. The electrolyte conducts ions between the electrodes and across the separator, whilst insulating against electron flow.

In broad terms, supercapacitors fall within two classes: electrostatic double layer capacitors (EDLCs), and pseudocapacitors. The former utilise a Helmholtz double layer that exists at the interface between electrode and electrolyte for storage of electrostatic charge. Pseudocapacitors rely upon faradic charge transfer reactions (including redox reactions, electrosorption and intercalation). When used as an electrode material, polyaniline undergoes redox reactions, thus polyaniline-based capacitors form one sub-class of pseudocapacitor.

Macroscopic mats of carbon nanotubes (CNTs) can serve as EDLCs as well as the current collector of pseudocapacitors [3–5]. These CNT mats are manufactured in industrial volumes by floating catalyst chemical vapour deposition (FCCVD) [6,7]; they are electrically conductive [8], possess an in-plane Young's modulus between 0.1 GPa and 10 GPa [9], a thermal conductivity between 20 W m$^{-1}$ K$^{-1}$ and 800 W m$^{-1}$ K$^{-1}$ [9], and a specific surface area in the range $1\times10^5$ m$^2$ kg$^{-1}$ to $4.8\times10^5$ m$^2$ kg$^{-1}$ [10–14]. Their favourable electrical, mechanical and thermal properties, and electrochemical stability, have led to research interest in their application as multifunctional current collectors for flexible [15] or stretchable [16] electronic products and for structural energy storage devices [17–19]. The CNT mat microstructure comprises an interconnected network of CNT bundles. Porous electrodes in pseudocapacitors are manufactured by first coating the CNT bundle network with a suitable electrode material, and second by immersing the porous electrodes into a liquid electrolyte [20–27]. In common with many other electrode architectures [28–31], the specific power of these CNT-based electrodes arises from a high internal surface area over which ionic transport and chemical reactions occur.

Recent interest in polyaniline (PANI), polyacetylene and polypyrrole as conducting polymers has arisen due to their high specific capacitance (exceeding $1 \times 10^6$ F kg$^{-1}$ [32,33]), low cost, and environmental stability [34]. The molecular structure of PANI evolves during charge and discharge, and exists in 3 idealised states in the base or salt forms: *Leucoemeraldine* base (L) or salt (LS), *Emeraldine* base (E) or salt (ES) and *Pernigraniline* base (P) or salt (PS). These idealised states are illustrated in Figure 1(b). PANI switches between base and salt forms via the doping/dedoping of H$^+$ depending on the pH of the electrolyte [34–36]. The proportion of N=C double bonds in the PANI molecule varies as a function of its state of charge due to reversible faradic reactions, which store and release electrical energy. For instance, PANI-(PS) is converted to PANI-(ES) or PANI-(LS), as protons (H$^+$) from the electrolyte and electrons (e$^-$) from the current collector react with the PANI molecule and convert covalent N=C double bonds into N-C plus N-H covalent bonds, see Figure 1(b). Upon conversion from the (L) state to the (P) state, the molecular unit gains a charge $\Delta Q_u = 4e$, where $e = 1.602 \times 10^{-19}$ C is the electron charge. The pseudocapacitance $C_u$ of the repeating unit is related to the charge transferred $\Delta Q_u$ and change in electrical potential during charging $\Delta \Phi$ according to $C_u = \Delta Q_u / \Delta \Phi$. As each repeating unit comprises four aniline molecules, its mass $m_u$ equals $4M_A/N_A$, where the molar mass of aniline is $M_A = 0.091$ kg mol$^{-1}$ [32] and $N_A = 6.022 \times 10^{23}$ mol$^{-1}$ is Avogadro's constant. Upon assuming that all PANI molecules are converted from the (L) state to the (P) state over the potential window $\Delta \Phi = 1.0$ V used herein, the theoretical maximum specific capacitance of PANI molecules $C_P = C_u/m_u$ equals $1.06 \times 10^6$ F kg$^{-1}$.

## 1.1 CNT-PANI composite electrodes

CNT-PANI composite electrodes exist in three morphological classes (i) to (iii), see Figure 2(a).

*Class (i): Short CNTs in powder form* are dispersed in a solution of aniline in acid, typically with sonication. The CNT-aniline solutions are polymerised to form CNT-PANI composites [20–22,37]. Alternatively, unidirectional CNT-PANI fibres are formed by injection of the CNT-PANI solutions into a bath of coagulating fluid [38].

*Class (ii): Aligned CNT forests* are manufactured upon substrates via a chemical vapour deposition (CVD) process. Subsequently, the forests are drawn into CNT-PANI composites directly via electrodeposition [39]. Alternatively, the forests may be processed into so-called 'forest-spun' mats or fibres in a dry state, prior to electrodeposition of a PANI layer [40–42].

*Class (iii): CNT mats or fibres* are formed from a CNT aerogel. The aerogel nucleates and grows within a floating catalyst chemical vapour deposition (FCCVD) reactor. The aerogel is then drawn continuously onto a rotating drum to form a 'direct-spun' fibre or mat, and can be coated with PANI via an electrodeposition process [17,23,43].

The tensile strength of CNT fibres increases with increasing CNT length [44] for all three classes. The tensile strength of fibres in class (ii) and class (iii) can exceed 1 GPa, which is much above the measured strength of class (i) materials [45]. Also the production volumes of materials manufactured via floating catalyst methods (classes (i) and (iii)) greatly exceed those of class (ii) [3,46]. For these reasons, composite electrodes of direct-spun CNT mats and PANI are promising candidates for structural energy storage devices.

## 1.2 Electrical performance of CNT-PANI capacitors

The reported specific power $P_s$ (the power per unit mass of the active material) and specific energy $W_s$ (the energy per unit mass of the active material) of the CNT-PANI composite electrode classes, as reported in the literature [17,20–24,37,40,41,43,47,48], and as measured in this study, are assembled in Figure 2(b), alongside data from electrolytic capacitors [49], EDLCs [21,47,50], PANI pseudocapacitors [24,26,32,47,51,52] and lithium-ion batteries [53]. Batteries offer higher specific energy than electrolytic capacitors, but possess much lower specific power. The specific energy and specific power of CNT-PANI electrodes span several orders of magnitude; their specific power overlaps with that of electrolytic capacitors and EDLCs, and their specific energy overlaps that of lithium ion batteries [23]. It is observed that the capacitance of micrometre-thick PANI films can decrease to only 25% of their theoretical maximum after only a few charge/discharge cycles [54]. In contrast, PANI films with thickness on the order of tens of nanometres are more resistant to such degradation, and exhibit much higher specific powers. Such findings motivate the use of PANI-coated direct-spun CNT mats as electrodes in a supercapacitor.

CNT-PANI composites comprise three distinct phases: CNT bundles, PANI and intervening voids, of volume fraction $f_B$, $f_P$ and $f_a$, respectively. The specific capacitance $C_s$ of the CNT-PANI composite is obtained by normalising the capacitance of the electrode by the PANI mass, and is plotted against the PANI volume fraction $f_P$ in Figure 2(c). CNT-PANI composites manufactured from CNT-PANI dispersions possess a CNT bundle volume fraction in the range $0.01 \leq f_B \leq 0.12$, whereas direct-spun CNT-PANI composites possess $0.18 \leq f_B \leq 0.29$, and forest-derived CNT-PANI composites have $0.28 \leq f_B \leq 0.49$. Data for CNT EDLCs are included in Figure 2(c), for which the specific capacitance $C_s$ is defined by electrode capacitance normalised by the mass of CNT bundles. The volume fraction of CNT bundles within CNT EDLCs is in the range $0.08 \leq f_B \leq 0.19$. Values for the specific capacitance of each CNT capacitor class plotted in Figure 2(c) are obtained at scan rates between 1 mV s$^{-1}$ and 200 mV s$^{-1}$ [17,20–24,26,32,37,40,41,43,47,48,51,52]. CNT-based EDLCs possess a specific power of up to 40 kW kg$^{-1}$; this value is close to the maximum specific power recorded for CNT-PANI electrodes. In contrast, the specific capacitance of all EDLCs is below $2 \times 10^5$ F kg$^{-1}$, a value which is exceeded by the specific capacitance of all CNT-PANI composite electrode classes [55]. The specific capacitance of the CNT-PANI composite electrodes varies widely, and depends upon electrode composition and scan rate of the voltage applied during characterisation with cyclic voltammetry. In broad terms, a lower scan rate and lower value of $f_P$ result in higher values of specific capacitance, and the variation in reported values of specific capacitance increases with increasing PANI volume fraction. Recall that the theoretical specific capacitance of PANI, as calculated above, is $C_P = 1.06 \times 10^6$ F kg$^{-1}$. This theoretical limit is included in Figure 2(c), and lies above the specific capacitance of all PANI-CNT electrodes as measured in this study and as summarised from the literature.

Studies of CNT-PANI composite electrodes have reported specific capacitances as high as $8 \times 10^5$ F kg$^{-1}$ [40], exceeding the performance of many other capacitor materials. Additionally, a tensile strength of up to 135 MPa and Young's modulus of 5 GPa makes them promising candidate materials for structural energy storage devices [17,38]. The successful commercial exploitation of these electrodes is hindered by the cyclic stability of CNT-PANI electrodes, which is currently inferior to that of other electrode materials [21,56]. This degradation arises due to side reactions during the oxidation and reduction of PANI molecules during charging and discharging, which reduce the molecular weight of the PANI [57]. Whilst the electrochemical behaviour of polyaniline has been examined extensively [17,20–23,37,40,41,43], only limited studies have been performed to date on the mechanical properties of coagulant-spun CNT-PANI fibres, and composite CNT-PANI electrodes, manufactured from CNT mats [17,38]. The influence of charge and discharge rate, electrode composition and state of charge upon the mechanical and electrochemical properties of these novel electrode composites is currently unclear, as are the mechanisms of degradation upon cycling and pre-charging. These topics are addressed in the present study.

**1.3 Scope of study**

CNT-PANI composites are manufactured by the electrodeposition of PANI onto direct-spun CNT mat. The composition of the composites is altered by varying the PANI electrodeposition time. The CNT volume fraction $f_B$ ranges from 0.19 to 0.29, and the volume fraction of PANI $f_P$ from 0.01 to 0.29. The CNT-PANI composites are porous, with void volume fraction $f_a$ in the range $f_a = 0.41$ to 0.78. The microstructure of the manufactured CNT-PANI composite electrodes is characterised with scanning electron microscopy. In-plane tensile tests are conducted to measure their stress-strain response, and the electrochemical properties are determined with cyclic voltammetry. The relationship between electrode microstructure and properties is explained by a micromechanical model.

# 2. Manufacture, composition and microstructure of CNT-PANI composites

## 2.1 CNT mat manufacture

The CNT mats[1] used in experiment were made by a productionised version of the floating-catalyst chemical vapour deposition (FCCVD) process of Li *et al.* [6]; the method is described in depth elsewhere [58]. The CNT mat, of width 0.9 m and thickness 70 μm, comprises many layers of CNT aerogel "sock", and is produced by drawing the sock continuously from an FCCVD reactor, and laying the sock flat upon a reel [9]. Each flattened sock is of width 80 mm and thickness 170 nm. We emphasise that each sock layer consists of an interconnected network of branched CNT bundles [8,9].

The macroscopic electrical conductivity, strength and modulus of direct-spun mats exhibit a varying degree of in-plane anisotropy. This anisotropy varies with the ratio of the drawing speed to the speed of the gas within the FCCVD reactor [9]. The principal material orientation is defined as that in which the CNT aerogel sock is drawn from the reactor. In order to remove amorphous carbons and residual metal catalyst particles that are present within the CNT mat after manufacture [60], the CNT mat is heated in air at 400°C for 1 hour and is then submerged in an aqueous hydrochloric acid solution of mass concentration 36.5% for 24 hours. The mat is then rinsed three times using ethanol and de-ionized water, and dried under vacuum overnight. This process decreases the mass of the mat by approximately 8%, and reduces its thickness from 70 μm to 50 μm.

---

[1] Tortech Nano Fibers Ltd., Hanassi Herzog St., Koren Industrial park, Ma'alot Tarshiha, 24952 Israel.

## 2.2 CNT-PANI composite manufacture

The electrodeposition procedure for PANI on CNT mat electrodes is sketched in Figure 3(a); it is the typical manufacturing route for CNT-PANI electrodes [21,23]. A CNT mat sample of width 10 mm, length 50 mm and 50 μm thickness was used as the working electrode. A platinum mesh[2] of width 50 mm, length 50 mm and thickness 0.5 mm was employed as a counter electrode, whilst a saturated calomel electrode[3] was used to provide a reference potential. The aqueous electrodeposition solution comprised 1 M sulphuric acid and 0.05 M aniline[4]. The CNT mat samples were coated with PANI by electrodeposition at a constant current of 5 mA with a potentiostat[5]; the deposition time ranged from 5 minutes to 60 minutes. The electrodeposition current was sufficiently low to ensure that a PANI layer of uniform thickness was deposited onto the CNT bundles. After electrodeposition, the CNT-PANI composite electrodes were thoroughly rinsed in ethanol and in de-ionised water three times; a neutral pH was confirmed with indicator paper. This washing step neutralises the coated PANI, reducing it to its base forms. The CNT-PANI electrodes were then dried under vacuum for 12 hours.

---

[2] Fisher Scientific U.K. Ltd., Bishop Meadow Road, Loughborough, LE11 5RG, UK.
[3] Scientific Laboratory Supplies Ltd., Wilford Industrial Estate, Ruddington Ln., Nottingham, NG11 7EP, UK.
[4] Sigma-Aldrich Co. Ltd., The Old Brickyard, New Road, Gillingham, SP8 4XT, UK.
[5] BioLogic SAS, 4 rue de Vaucanson, Seyssinet-Pariset 38170, France.

## 2.3 Composition and microstructure of CNT-PANI composite electrodes

The volume fractions $f_B$ of CNT bundles and $f_P$ of PANI relate to the corresponding mass fractions $m_B$ and $m_P$ as follows. Define $\rho_B = 1560$ kg m$^{-3}$ as the measured CNT bundle density from [9] and $\rho_P = 1245$ kg m$^{-3}$ as the density of PANI in the mixed state from [61]. Upon denoting $\rho$ as the measured density of the CNT-PANI composes, it follows that

$$f_B = m_B \frac{\rho}{\rho_B}, \qquad f_P = m_P \frac{\rho}{\rho_P}. \qquad (1)$$

The mass fraction $m_B$ is the ratio of CNT bundle mass to composite mass; consequently the PANI mass fraction is $m_P = 1 - m_B$. A mass balance was used to determine the mass of the CNT mat samples prior to electrodeposition and that of the CNT-PANI composites after manufacture. Upon taking the CNT mat mass in each electrode to equal that measured before electrodeposition, the PANI mass in each electrode is obtained by subtraction of the CNT mat mass from the electrode mass as measured after manufacture. A Vernier scale was used to measure the sample width and length whilst the thickness was measured using a micrometer. The inferred densities of the CNT mat and CNT-PANI composites are recorded in Table 1, along with the volume fractions of CNT bundles $f_B$, PANI $f_P$ and voids $f_a$. Samples are labelled as CNT-PANI-T$x$, where $x$ denotes the deposition time in minutes, ranging from 5 minutes to 60 minutes. As-manufactured PANI exists in a mixed state of (P), (E) and (L). The mass fraction of PANI increases with increasing deposition time, see Figure 3(b).

Images of the microstructure of the CNT-PANI electrodes and the CNT mat were obtained with a scanning electron microscope[6] (SEM), using a 3 μm spot size and 5 kV acceleration voltage. Images of the CNT mat and CNT-PANI electrode microstructure are given in Figure 3(c)-(h). The CNT mat microstructure, as shown in Figure 3(c), reveals a random CNT bundle network of branched CNT bundles. In composite electrodes of low PANI content, the PANI forms a uniform coating upon the surface of the bundles, see Figure 3(d). As the PANI content increases, the voids between the CNT bundles are filled, see Figure 3(e)-(h). Electrolyte can flow through the interconnected porosity upon immersion. In the resulting pseudocapacitor, the PANI coating forms the active material, and the CNT bundle network serves as the current collector.

---
[6] Carl Zeiss Ltd., Zeiss House, Building 1030, Cambourne Business Park, Cambourne, CB23 6DW, UK.

The multi-walled and branched nature of the CNT mat was confirmed by a preliminary study using a transmission electron microscope (TEM). Individual layers of the CNT mat were placed upon a copper mesh and imaged with field emission TEM[7], to reveal the microstructure of the CNT bundles, using an acceleration voltage of 120 kV and spot size of 2 nm. The images, shown in Figure S1 of the supplementary information, confirm that the CNTs branch between bundles, forming an interconnected network. The CNTs possess multiple walls, and are of typical diameter between 10 nm and 15 nm.

The elemental composition of the CNT mat and CNT-PANI-T20 electrode were determined by collecting energy-dispersive X-ray spectra from the mat, using a 20 kV electron beam at 8.5 mm working distance, and a suitable probe[8]. The spectra are show in Figure S2(a) of the supplementary information. Both samples are predominantly carbon; additional peaks measured for the CNT mat correspond to residual iron oxide and sulphur catalyst particles; traces of chlorine are also present due to acid washing. For the CNT-PANI-T20 electrodes, additional sulphur is imparted by the electrolyte, and a minor peak for nitrogen is also present, see Figure S2(b). This peak corresponds to the nitrogen atoms within PANI. Gold peaks present in both spectra are from nanometre-thick gold coatings applied to enable sufficient electrical conductivity for imaging. Elemental maps for the elements carbon, iron and sulphur are given in Figure S3 of the supplementary information, and correspond to the accompanying image of CNT mat microstructure obtained with a scanning electron microscope. Carbon is present throughout the mat microstructure, as are the residual sulphur and iron catalyst particles.

---

[7] FEI Tenaci F20 field emission electron microscope. 5350 NE Dawson Creek Drive, Hillsboro, Oregon, 97124, USA.
[8] Oxford Instruments, Tubney Wood, Abingdon, OX13 5QX, UK.

The crystal structure of the CNT mat and CNT-PANI composite electrodes were investigated with an X-ray diffractometer[9]; a copper source provided radiation of wavelength 1.54 Å. Spectra recorded for a CNT mat sample and CNT-PANI-T20 electrode are given in Figure S4 of the supporting information. The peaks within both spectra at 26°, 44° and 54° are typical for multi-walled carbon nanotubes and other graphitic structures [62]. The addition of polyaniline results in no discernible alteration to the spectra, confirming that the deposited PANI is amorphous, and does not modify the crystal structure of CNTs within the mat.

# 3. Electrical and electrochemical properties of as-manufactured CNT-PANI composite electrodes

## 3.1 Electrical conductivity of CNT-PANI composite electrodes

The planar electrical conductivity of CNT mat samples and CNT-PANI composite electrodes were determined in the dry state (prior to infiltration by the electrolyte) with a direct-current four-point probe method. The sample dimensions and the experimental method are all given in Figure S5(a) within the supplementary information. Electrical contacts with negligible resistance were made by laying the samples upon copper wires.

The electrical conductivity $K$ measured parallel to the principal material direction for CNT mat and CNT-PANI composite electrode samples is plotted as a function of the CNT bundle volume fraction $f_B$ in Figure 4(a); $K$ varies from 180 kS m$^{-1}$ to 310 kS m$^{-1}$ depending upon composition. Since the electrical conductivity of PANI is between 10$^{-3}$ kS m$^{-1}$ and 10 kS m$^{-1}$ [63,64], the electrical conductivity of the CNT-PANI composites is dominated by the contribution from the CNTs, such that $K$ scales linearly with $f_B$; the contribution to $K$ from the PANI is negligible. For completeness, in-plane measurements of the electrical conductivity were made for dry specimens of CNT mat and for the composite CNT-PANI-T60 electrode oriented at 90° to the principal material orientation. These measurements are included in Figure 4(a). The electrical conductivity measured at 90° to the principal material direction is about 67% of that measured in the principal material direction, and is dependent upon the CNT bundle volume fraction. The CNT mat and CNT-PANI-T60 composites both exhibit similar in-plane anisotropy in electrical conductivity.

---

[9] Siemens D500 X-ray diffractometer. Brucker UK Ltd., Banner lane, Coventry, CV4 9GH. UK.

## 3.2 Electrochemical properties of CNT-PANI composite electrodes

The three-electrode system, as illustrated in Figure 3(a), was also used to characterise the electrochemical properties of CNT-PANI electrodes. The working electrode consisted of the CNT-PANI composites; the counter electrode consisted of platinum mesh, and the saturated calomel electrode was used to provide a reference potential. The potentials of the reference, counter and working electrodes are denoted $\Phi_1$, $\Phi_2$ and $\Phi_3$ respectively, see Figure 3(a). The measured reference electrode potential is corrected so that the potential $\Phi_1$ is that of the standard hydrogen electrode throughout. An aqueous 2 M sulphuric acid solution was used as the electrolyte (instead of the electrodeposition medium of 1 M sulphuric acid and 0.05 M aniline).

Consider the electrochemical characterisation of the CNT-PANI electrodes upon charging and discharging. After immersion into the electrolyte solution, samples were allowed to equilibrate for a period of five minutes with no externally applied current. The open circuit voltage (OCV), denoted by $\Phi_{OCV}$ was 0.43 V, regardless of the mass of PANI deposited. (This implies that the state of as-deposited PANI is insensitive to PANI mass.) Cyclic voltammetry tests were conducted by applying a triangular voltage waveform $\Phi = \Phi_3 - \Phi_1$ of period $T$, as sketched in Figure 4(b), and by measuring the resultant current flow $I(t)$. The state of charge $Q(t)$ is the integral of the current flow with respect to time. The voltage $\Phi$ was cycled between a maximum value $\Phi_{max} = 0.8$ V and a minimum value $\Phi_{min} = -0.2$ V in order to avoid undesired reactions [65,66], and the scan rate $\dot{\Phi} = d\Phi/dt$ was varied between 1 mV s$^{-1}$ and 50 mV s$^{-1}$.

The measured current $I(t)$ and resulting charge $Q(t)$ from cyclic voltammetry of the CNT-PANI electrodes are plotted against voltage $\Phi(t)$ during cycling at scan rate $\dot{\Phi} = 1$ mV s$^{-1}$ in Figure 4(c) and Figure 4(d), respectively. Here, time $t$ is an implicit parameter. Recall that during charge and discharge, PANI molecules convert between the (LS), (ES) and (PS) states, see Figure 1(b). A degradation product, p-aminophenol, is formed at the end of the PANI molecule chains, and also undergoes oxidation and reduction during electrochemical cycling [57,67]. The pair of redox peaks, as seen in the cyclic voltammetry response at $-0.2 \leq \Phi \leq 0.3$ V in Figure 4(c) correspond to the conversion of PANI molecules between the (LS) and (ES) states (see Figure 1(b)); in contrast the pair of peaks at $0.4 \leq \Phi \leq 0.6$ V correspond to oxidation and reduction of degradation products [57,67,68]. For a chosen scan rate, the rate of oxidation, governed by the egress of protons from the PANI, differs from the rate of reduction of the PANI upon discharge [54]. Because of this mismatch in oxidation and reduction rates, charge is accumulated during the first voltage cycle for CNT-PANI electrodes, see Figure 4(d).

The electrode capacitance $C$ is calculated from the first cycle of the charge versus voltage response plotted in Figure 4(d). $C$ is related to the voltage drop during discharge $\Phi_{max} - \Phi_{min}$ and to the net charge transferred between the beginning and end of the voltage discharge ramp $\Delta Q = Q(\Phi_{max}) - Q(\Phi_{min})$, according to $C = \Delta Q/(\Phi_{max} - \Phi_{min})$. The electrode capacitance is plotted against the mass of PANI within the electrode $M_P$ in Figure 4(e). Capacitance increases with increasing PANI mass, and increases with decreasing scan rate.

Now consider the conversion efficiency $\eta$ of the CNT-PANI electrodes, defined as the mass of PANI coating (on the CNT bundles) which is chemically converted between the (LS) and (PS) states as illustrated in Figure 1(b) upon charging and discharging, divided by the total measured mass of PANI in each electrode. The conversion efficiency $\eta$ is estimated as follows:

(i) Upon neglecting the capacitance of the double layer within the electrolyte [55], the mass of PANI converted during cycling $\Delta M_P$, is related to the molar mass of aniline, $M_A$, the Faraday constant $F = 9.649 \times 10^4$ C mol$^{-1}$, and the measured charge transferred during each charge/discharge cycle $\Delta Q$ according to

$$\Delta M_P = \frac{\Delta Q}{F} M_A \qquad (2)$$

(ii) The conversion efficiency is the ratio of mass of PANI that is converted during cycling $\Delta M_P$ to the total measured mass of PANI in each electrode $M_P$, such that

$$\eta = \frac{\Delta M_P}{M_P}. \qquad (3)$$

The conversion efficiency $\eta$ is plotted against PANI volume fraction $f_p$ in Figure 4(f), for PANI volume fractions in the range $0.015 \leq f_P \leq 0.294$. The conversion efficiency of CNT-PANI electrodes increases with decreasing scan rate and with decreasing PANI volume fraction, due to the time-dependence of the Faradic reactions upon charge and discharge [55,63]. A maximum conversion efficiency of 0.63 is achieved for a PANI volume fraction $f_P = 0.015$ and a scan rate of $\dot{\Phi} = 1.0$ mV s$^{-1}$.

# 4. Mechanical properties of as-manufactured CNT-PANI composite electrodes

A screw-driven test machine was employed to measure the uniaxial stress-strain response of CNT-PANI composites and the CNT mat at strain rate $\dot{\epsilon} = 10^{-4}$ s$^{-1}$. Samples of mat and composite electrode were of length 50 mm and width 10 mm. The degree of in-plane anisotropy was assessed by conducting tensile tests upon samples of CNT mat and CNT-PANI-T60 composite with the loading direction at angles of 0° and 90° relative to the principal material direction. Specimen ends were held between paper end-tabs with wedge grips, see Figure S5(b). To measure the strain state within the sample, two 0.5 mm diameter dot stickers were applied to the sample with spacing of 25 mm; their relative displacement was measured throughout the test with a video camera[10] at a frequency of 1 Hz. The yield strength of the direct-spun CNT mat is deduced from its nominal stress versus strain response using a bilinear fit, whereas the CNT-PANI composite electrode yield strength is obtained from the intersection between their stress-strain response and a line of slope identical to that of the initial, linear part of the stress-strain curve but offset in strain by 0.2% from the origin.

Tensile nominal stress $\sigma$ versus nominal strain $\epsilon$ responses of CNT mat and CNT-PANI composite electrodes are plotted in Figure 5(a) for straining along the principal material direction. The effect of material orientation is demonstrated in Figure 5(b), in which the stress versus strain response of the CNT mat and CNT-PANI-T60 electrode are plotted for samples tested at 0° and at 90° to the principal material direction. Both the CNT mat and the electrodes materials exhibit moderate in-plane anisotropy. The tensile stress-strain responses of the CNT-PANI electrodes lie much above those of the CNT mat. The modulus $E$ and yield strength $\sigma_{YS}$ of both CNT-PANI electrodes and CNT mat are plotted against electrode density $\rho$ in Figure 5(c) and Figure 5(d), respectively: both modulus and strength increase with increasing electrode density.

---

[10] GOM UK Ltd., No. 14 Cobalt Centre, Siskin Parkway East, Coventry, West Midlands, CV3 4PE, UK.

# 5. Effects of pre-charging and cycling upon the composition, electrochemical and mechanical properties of electrodes

Structural capacitors ideally possess a stable charge capacity, a high resistance to capacity fade, and a small drop in mechanical properties after multiple charge/discharge cycles [19]. In order to assess the promise of CNT-PANI electrodes for engineering applications, the effect of pre-charging and multiple charge/discharge cycles upon their composition, electrochemical and mechanical properties are now determined.

## 5.1 Pre-charging and cycling of CNT-PANI electrodes

CNT-PANI-T20 electrodes were either pre-charged or cycled using the three-electrode system illustrated in Figure 3(a).

*Pre-charging:* Electrodes were pre-charged by monotonically increasing or decreasing the potential $\Phi$ at a constant ramp rate of $\dot{\Phi} = \pm 1$ mV s⁻¹, until a specified potential difference $\Phi_{PRE}$ was reached, see Figure 6(a). The voltage was then held constant at the chosen value of $\Phi_{PRE}$ for a period of five minutes.

*Cycling:* In a separate experiment, pristine CNT-PANI-T20 electrode samples were subjected to 1000 charge-discharge cycles with the voltage waveform of Figure 4(b) at scan rate $\dot{\Phi} = 10$ mV s⁻¹, such that $\Phi_{max} = 0.8$ V and $\Phi_{min} = -0.2$ V. In order to determine the effect of electrochemical cycling upon the CNT bundle network, a sample of CNT mat absent PANI (T0) was also subjected to 1000 charge/discharge cycles.

After either pre-charging or cycling, the power supply was disconnected, the samples were rinsed in ethanol and then water three times as described above to reduce the PANI to its base forms, and then dried under vacuum. The composition and the electrochemical, electrical and mechanical properties of the pre-charged and cycled electrodes were then characterised. The details are as follows.

## 5.2 Effect of pre-charging and cycling upon CNT-PANI capacitance and electrode composition

*CNT-PANI capacitance*

The electrochemical properties of electrodes that were first subjected to pre-charging or were first subjected to 1000 charge/discharge cycles were investigated with cyclic voltammetry tests, as described in section 3.2. The measured current $I(t)$ and charge $Q(t)$ recorded during cyclic voltammetry tests on CNT-PANI-T20 electrodes post pre-charging and post cycling are plotted against applied voltage $\Phi(t)$ in Figure 6(b) and Figure 6(c) respectively. The cyclic responses of a CNT-PANI-T20 electrode, and a CNT mat electrode absent PANI, are included for comparison: cyclic voltammetry tests were performed immediately after manufacture for these electrodes (that is, without pre-charging or cycling).

For the electrodes subjected to pre-charging or cycling, the current measured during cyclic voltammetry and plotted in Figure 6(b) is below that measured for the CNT-PANI-T20 electrode immediately after manufacture. The reduction in measured current and charge transferred during cycling, compared to that measured immediately after manufacture, is greatest for electrodes pre-charged to the highest voltage $\Phi_{PRE} = 1$ V.

The capacitance of the CNT mat and CNT-PANI-T20 electrodes measured over the course of 1000 charge-discharge cycles is plotted in Figure S6(a) of the supplementary information. Measurements of electrode capacitance are normalised to their values at the start of the experiment. The change in the capacitance of the CNT mat electrode is minor. The capacitance of the CNT-PANI-T20 electrode decreased by less than 15% of its initial value over the application of 1000 charge-discharge cycles. Most of this decrease occurred over the first 500 cycles.

*Raman spectroscopy on pre-charged specimens*

The electrodeposited PANI layer upon CNT bundles, in the as-manufactured state, comprises PANI co-existing in all three of its chemical states (P), (E) and (L). Recall from Figure 1(b) that the proportion of N=C double bonds in the PANI molecule varies for these 3 chemical states; also, charge and discharge alters the proportion of N=C double bonds and thereby alters the relative proportion of these 3 states.

The density of N=C double bonds has been assessed via Raman spectroscopy as follows. Raman spectra were obtained for the CNT-PANI-T20 electrodes after pre-charging, washing, rinsing and drying with a spectrometer[11], using a 532 nm wavelength, 50x microscope objective, and a spot size of 1.3 μm. These spectra are presented in Figure S7 of the supplementary information; the signal intensity is normalised by the height of the highest peak within the spectra at 158 mm$^{-1}$. A Raman spectrum is included for a CNT mat electrode with PANI absent. The spectra confirm that there is change in the number of N=C double bonds after pre-charging to different voltages, as reported elsewhere [57].

*CNT-PANI electrode composition*

Now consider the effect of pre-charging or cycling upon electrode composition. The volume fractions of PANI, CNT bundles and of porosity for pre-charged and cycled CNT-PANI-T20 electrodes were measured as described in section 2.3; results are recorded in Table 2. Recall that all samples are manufactured from CNT mat samples of identical mass.

---

[11] Horiba UK Ltd., Kyoto Close, Moulton Park, Northampton, NN3 6FL, UK.

The mass and dimensions of the CNT mat electrodes absent PANI were insensitive to electrochemical cycling, and thus the mass of CNT bundles within the CNT-PANI electrodes was taken to be constant regardless of the electrode charging history. For CNT-PANI electrodes, the action of pre-charging or application of 1000 charge/discharge cycles swells the electrode in the through-thickness direction. This swelling reduces the PANI volume fraction $f_P$ and the CNT bundle volume fraction $f_B$. For the samples measured herein, the PANI volume fraction $f_P$ decreased by the largest amount (from 0.087 to 0.015) for electrodes pre-charged to $\Phi_{PRE} = 1.0$ V, due to a 31% increase in electrode thickness.

The capacitance of cycled and pre-charged CNT-PANI-T20 electrodes is plotted in Figure 6(d).as a function of mass fraction of PANI $m_P$, as measured after cycling or pre-charging. All electrodes subjected to pre-charging or cycling possess a PANI mass fraction $m_P$ below that measured immediately after manufacture. For completeness, electrode capacitance and PANI mass fractions measured immediately after manufacture for other CNT-PANI electrodes with $0 \leq m_P \leq 0.25$ are included in Figure 6(d). Regardless of initial composition or charge/discharge history, the electrode capacitance varies almost linearly with the measured PANI mass fraction $m_P$. Reasons for the reduction in PANI mass fraction $m_P$ upon pre-charging or cycling, and reasons for the invariant relationship between $m_P$ and electrode capacitance over the charge/discharge history are now discussed with reference to the microstructure of the electrodes subjected to pre-charging.

Images of the CNT-PANI-T20 electrode microstructure after pre-charging are given in Figures 6(e) and (f). Consider first the microstructure of CNT-PANI-T20 electrodes pre-charged to a voltage $\Phi_{PRE} = 1.0$ V, see Figure 6(e). The quantity of PANI visible upon the surface of the CNT bundles after pre-charging to a voltage $\Phi_{PRE} = 1.0$ V is much below that observed immediately after manufacture in Figure 3(f), and confirms that pre-charging to $\Phi_{PRE} = 1.0$ V removes PANI from the surface of the CNT bundle network. The mechanism by which PANI is removed from the surface of the CNT bundles at such voltages has been studied elsewhere [30], and is attributed to the formation of soluble degradation products due to reactions at the ends of the PANI molecular chains. Upon pre-charging to a voltage $\Phi_{PRE} = 1.0$ V gas bubbles were visible upon the electrode surface as hydrolysis reactions took place, and a dark green solution of degradation products enveloped the electrode. However, upon rinsing, washing and drying the soluble degradation products were largely removed from the electrode surface, leaving the remaining PANI upon the surface of the CNT bundle network.

Recall that the swelling upon pre-charging to $\Phi_{PRE} = -0.2$ V reduces the CNT bundle volume fraction $f_B$ by 14%, see Table 2. An SEM image of the CNT-PANI-T20 electrode microstructure after pre-charging to a voltage $\Phi_{PRE} = -0.2$ V is given in Figure 6(f). Sites of non-uniform swelling are visible on the surface of the PANI coating. The electrode swells in the out-of-plane direction during reduction as H$^+$ ions from the electrolyte ingress into, and bond onto, the PANI layer [69], see Figure 1(b). An image of the microstructure of the CNT-PANI-T20 electrode obtained after 1000 cycles is given in Figure S6(b) of the supplementary information. It is deduced that the microstructure of the cycled electrode is little different from that of the CNT-PANI-T20 electrode imaged after manufacture.

*Dependence of capacitance upon PANI mass fraction*

Now consider the relationship between PANI mass fraction $m_P$ and electrode capacitance $C$. Upon electrochemical cycling, degradation products are formed at the end of the PANI molecules [30]. Pre-charging or cycling leads to such a degradation and these soluble degradation products are removed by rinsing and drying of the electrodes before the electrode composition and capacitance is measured [57]. The PANI that remains upon the CNT bundles is free from degradation products, and consequently the cycled or pre-charged electrodes possess a capacitance equal to that of an electrode of similar composition measured immediately after manufacture. This explains the invariant relationship between PANI mass fraction and electrode capacitance, regardless of charge history.

## 5.3 The tensile stress-strain response of pre-charged and cycled CNT-PANI electrodes

We proceed to report the effect of PANI state of charge, or charge-discharge history, upon the stress-strain response of CNT-PANI composite electrodes. The stress-strain responses of CNT-PANI-T20 electrode samples, as measured in the dry state after charging with the 3-electrode system and potential ramps of Figure 6(a), are plotted alongside the stress-strain response of the as-manufactured CNT-PANI electrode in Figure 7(a). The measured stress-strain response of samples pre-charged to $\Phi_{PRE} = 0.1$ V, $\Phi_{PRE} = 0.3$ V and $\Phi_{PRE} = 0.6$ V all lie close to that of the as-manufactured electrode. In contrast, the modulus and yield strength of electrodes pre-charged to the more extreme values of $\Phi_{PRE} = 1.0$ V or $\Phi_{PRE} = -0.2$ V are much below the corresponding values for the as-manufactured electrode.

The uniaxial stress strain response of a CNT-PANI-T20 electrode and a CNT mat electrode absent PANI are compared in Figure 7(b); in each case, the effect of 1000 charge/discharge cycles upon the response is determined. Pre-cycling degrades the modulus and strength of the CNT-PANI-T20 electrode but has only a negligible effect upon the mechanical response of the CNT mat.

## 6. A micromechanical model for the tensile stress-strain response of CNT-PANI composite electrodes

A micromechanical model is now constructed to obtain insight into the dependence of the stress-strain response of CNT-PANI electrode composites upon composition, and to understand the effect of cycling and pre-charging upon the stress-strain response. The model takes as input the mechanical response of the underlying CNT, the observed microstructure of the CNT-PANI electrodes, and the mechanical properties of PANI as reported in the literature [70,71].

## 6.1 An idealised CNT-PANI composite electrode microstructure

Recall that the CNT-PANI composite electrodes reported herein comprise an electrodeposited PANI coating upon a network of interconnected CNT bundles that constitute the mat. It is known from a previous study [9] that, under uniaxial tension, dry CNT mats deform in the manner of an open-cell foam: the CNT bundles bend and shear, but are resistant to axial stretch. The observed microstructure of the CNT-PANI composite electrodes prompts the choice of a periodic, hexagonal honeycomb planar unit cell, such that CNT struts of thickness $t_B$ are formed from interconnected CNT bundles, and a PANI layer of thickness $t_P$ coats the CNT struts, see Figure 8(a). This 2D unit cell has been used previously in a successful manner to predict the elastic-plastic stress-strain behaviour of CNT mat-epoxy composite materials [6]: the microstructure of this composite similarly comprises a polymer-coated CNT bundle network, but with PANI instead of epoxy.

The CNT bundle volume fraction, PANI volume fraction and void volume fraction within the unit cell are assigned the values recorded in experiment and written in Tables 1 and 2. The PANI coats the CNT bundle struts and partially fills the voids, see Figure 8(a). All struts within the honeycomb are of length $l$ and of width $t_B$; the fillet radii $R_1$ and $R_2$ between the joined CNT bundle struts and bordering the hexagonal pores are written in terms of the strut thickness $t_B$ such that $R_1 = 2.0 t_B$ and $R_2 = 0.54 t_B$.

The macroscopic stress versus strain response of the unit cell of Figure 8(b) was predicted with plane-stress finite element calculations, using the finite element software

ABAQUS[12]. Eight noded quadratic elements were used for the CNT bundles and PANI respectively; the mesh was suitably refined to ensure convergence of the simulation results, as shown in Figure S8. Details on the rate independent elasto-plastic response of the CNT bundles and PANI are given in the following section. Perfect bonding is assumed between the PANI and CNT bundles. Anisotropy is modelled by suitable choice of the angle $\omega$. This angle depends upon the degree of alignment of the CNT bundle network with the drawing of the mat during manufacture.

Symmetries of the 2D unit cell and of the uniaxial tensile loading along the vertical $x_1$ direction imply that the top of the unit cell is displaced by a uniform displacement $u_2$ whereas the rightmost side face of the unit cell is allowed to displace by a displacement $u_1$ such that the lateral net force vanishes. These periodic boundary conditions are illustrated in Figure 8(b).

## 6.2 CNT bundle and PANI constitutive models

The high axial stiffness and strength of CNT walls arise from the covalent bonding between carbon atoms. In contrast, the comparatively weak bonds between bundled CNTs lead to bundles of low longitudinal shear modulus and low shear yield strength [72–74]. Here, the CNT bundle is treated as an anisotropic, homogeneous continuum, with in-plane moduli $(E_{11}^B, E_{22}^B, E_{33}^B)$, shear moduli $(G_{23}^B, G_{12}^B, G_{12}^B)$, and Poisson ratios $(\nu_{12}^B, \nu_{21}^B, \nu_{13}^B, \nu_{31}^B \nu_{23}^B, \nu_{32}^B)$. Consider the elastic stress-strain behaviour of CNT bundles. Their elastic strain state $\varepsilon_{ij}^e$ relates to their stress state $\sigma_{ij}$ according to the following constitutive relationship:

---

[12] ABAQUS Standard (version 6.14). Dassault Systèmes Simulia Corp, 334 Cambridge Science Park, Milton Rd., CB3 0WN, Cambridge, UK.

$$\begin{Bmatrix} \varepsilon_{11}^e \\ \varepsilon_{22}^e \\ \varepsilon_{33}^e \\ \gamma_{23}^e \\ \gamma_{13}^e \\ \gamma_{12}^e \end{Bmatrix} = \begin{bmatrix} 1/E_{11}^B & -v_{21}^B/E_{22}^B & -v_{31}^B/E_{33}^B & 0 & 0 & 0 \\ -v_{12}^B/E_{11}^B & 1/E_{22}^B & -v_{32}^B/E_{33}^B & 0 & 0 & 0 \\ -v_{13}^B/E_{11}^B & -v_{23}^B/E_{22}^B & 1/E_{33}^B & 0 & 0 & 0 \\ 0 & 0 & 0 & 1/G_{23}^B & 0 & 0 \\ 0 & 0 & 0 & 0 & 1/G_{13}^B & 0 \\ 0 & 0 & 0 & 0 & 0 & 1/G_{12}^B \end{bmatrix} \begin{Bmatrix} \sigma_{11} \\ \sigma_{22} \\ \sigma_{33} \\ \tau_{23} \\ \tau_{13} \\ \tau_{12} \end{Bmatrix} \quad (4)$$

Estimates for the elastic constants in equation (4) were obtained in our earlier studies [8,9] and are used again here. The longitudinal bundle modulus is taken to be $E_{11}^B = 680$ GPa, the transverse modulus is $E_{22}^B = E_{33}^B = 50$ GPa, and Poisson ratio is $v_{12}^B = v_{13}^B = 0.3$. Since the much lower values of bundle shear moduli $G_{12}^B$, $G_{13}^B$ and $G_{23}^B$ arise from weak bonds between adjacent nanotubes, and are assumed equal. The value of longitudinal bundle shear modulus is $G_{12}^B = 9.5$ GPa [8].

The post-yield inelastic behaviour of CNT bundles is represented using the Hill anisotropic yield criterion [75]. This criterion defines the total strain rate $\dot{\varepsilon}_{ij}$ as a summation of the elastic and plastic strain rates $\dot{\varepsilon}_{ij}^e$ and $\dot{\varepsilon}_{ij}^p$,

$$\dot{\varepsilon}_{ij} = \dot{\varepsilon}_{ij}^e + \dot{\varepsilon}_{ij}^p. \quad (5)$$

The associated flow rule is based upon the plastic strain rate, the Hill potential $f(\sigma_{ij})$, and a plastic multiplier $\dot{\lambda}$, as follows:

$$\dot{\varepsilon}_{ij}^p = \dot{\lambda} \frac{\partial f}{\partial \sigma_{ij}}, \quad (6)$$

Here, the Hill potential $f$ is defined in terms of direct stresses $\sigma_{ij}$, shear stresses $\tau_{ij}$, and material constants $(F, G, H, L, M, N)$, as

$$2f = F(\sigma_{22} - \sigma_{33})^2 + G(\sigma_{33} - \sigma_{11})^2 + H(\sigma_{11} - \sigma_{22})^2 + 2L\tau_{23}^2 + 2M\tau_{31}^2 + 2N\tau_{12}^2 \quad (7)$$

The uniaxial tensile yield strengths of a CNT bundle in three orthogonal directions, as denoted by $\sigma_{11}^B$, $\sigma_{22}^B$ and $\sigma_{33}^B$, are used to define the constants $F$, $G$ and $H$ as follows:

$$G + H = \frac{1}{(\sigma_{11}^B)^2}; \qquad F + H = \frac{1}{(\sigma_{22}^B)^2}; \qquad G + F = \frac{1}{(\sigma_{33}^B)^2}. \tag{8}$$

The constants $L$, $M$ and $N$ follow from the bundle shear yield strengths $\tau_{ij}^B$, where

$$L = \frac{1}{2(\tau_{23}^B)^2}; \qquad M = \frac{1}{2(\tau_{31}^B)^2}; \qquad N = \frac{1}{2(\tau_{12}^B)^2}. \tag{9}$$

The tensile fracture strength of CNT walls is denoted $\sigma_w$; its value lies in the range 5.5 GPa to 25 GPa [76]. The axial yield strength of the CNT bundles is taken to be related to wall fracture strength $\sigma_w$ and bundle density $\rho_B$ according to $\sigma_{11}^B = \sigma_w(\rho_B/\rho_w)$. Upon taking $\sigma_w = 5.5$ GPa [76], it follows that $\sigma_{11}^B = 3.7$ GPa. The values of yield strength of CNT bundles in longitudinal shear ($\tau_{12}^B$, $\tau_{13}^B$), in transverse shear $\tau_{23}^B$, and the remaining transverse normal yield strengths $\sigma_{22}^B$ and $\sigma_{33}^B$, are set equal to $\tau_y^B$. To ensure convergence of numerical calculations, a hardening modulus of value $10^{-4} E_{11}^B$ was attributed to all stresses post-yield.

The PANI phase is taken to be isotropic, elastic-ideally plastic. The assumed PANI Young's modulus is $E^P = 1.4$ GPa, Poisson ratio is $\nu_P = 0.3$ and yield strength is $\sigma_Y^P = 83$ MPa, as reported in the literature [70,71]. Values for all material constants adopted in finite element computations are summarised in Table 3.

## 6.3 Calibration and prediction of the unit cell model for CNT mat and CNT-PANI electrodes after manufacture

The honeycomb model was calibrated as follows. First, the elastic modulus of the dry unit cell absent PANI was evaluated in directions $x_1$ and $x_2$ of Figure 8(b) for $30° \leq \omega \leq 50°$. The ratio of macroscopic moduli $E_{11}/E_{22}$ from the unit cell calculations is plotted against $\omega$ in Figure S9. The value $E_{11}/E_{22} = 1.6$ measured in experiment infers that $\omega = 34°$. Second, predictions were obtained for the uniaxial stress versus strain response of all CNT-PANI electrodes in the $x_1$ direction, and for the CNT and CNT-PANI-T60 electrodes in both $x_1$ and $x_2$ directions. In all simulations, the initial geometry of the unit cell was set with $\omega = 34°$, and the shear yield strength assigned to the CNT bundles $\tau_y$ was varied. Upon selection of the value $\tau_y^B = 480$ MPa, adequate agreement was obtained with the stress-strain response for both the CNT mat and all CNT-PANI electrode compositions measured after manufacture, see Figure 8(c).

## 6.4 Prediction of the calibrated model for samples subjected to pre-charging and cycling

Now consider the predictions of the above model for the CNT-PANI electrodes that were subjected to pre-charging at $\Phi_{\text{PRE}} = 1$ V and $\Phi_{\text{PRE}} = -0.2$ V or to 1000 charge/discharge cycles. The simulated stress-strain responses of pre-charged and cycled CNT-PANI-T20 electrodes, based upon the same unit cell model and CNT and PANI volume fractions as measured after pre-charging or cycling, are plotted in Figure 8(d) alongside that of the CNT-PANI-T20 electrode after manufacture. Predictions for the stress-strain response of the pre-charged and cycled samples also agree closely with the measurements.

The Young's modulus and tensile yield strength measured parallel with the principal direction of the CNT mat and all CNT-PANI electrodes are plotted as a function of the measured bundle volume faction in Figure 8(e) and 8(f), respectively. Data for electrodes subjected to 1000 charge/discharge cycles and pre-charging are included. Predictions for the electrode modulus and strength based upon the CNT bundle honeycomb model with PANI volume fraction $f_P = 0$ and $f_P = 0.3$ are included over the compositional range. In agreement with the experimental observations, the predicted effect of PANI content upon electrode modulus and strength is relatively minor compared to that of the CNT bundle volume fraction over the range of electrode compositions studied herein. The measured modulus and strength both lie close to values predicted by the honeycomb model. We conclude that the modulus and yield strength of the CNT-PANI electrodes after manufacture, after pre-charging or after cycling, is primarily dictated by the CNT bundle volume fraction.

## 7. Concluding remarks

The specific power, specific energy, and specific capacitance of CNT-PANI composites varies across multiple orders of magnitude. The performance of CNT-PANI electrodes is sensitive to the electrode composition and to the voltage scan rate. In the present study, composites of direct-spun CNT mats and polyaniline were manufactured by an electrodeposition technique. Control of the electrodeposition time allowed for a wide variation in composite composition: the CNT bundle volume fraction varied from 0.19 to 0.29, PANI volume fraction from 0 to 0.29, and the void volume fraction from 0.14 to 0.79. PANI adheres to CNT bundles within the mats, forming a thin layer at low volume fractions. With increased PANI volume fraction, the PANI layer extends into the porosity between CNT bundle struts.

The in-plane electrical conductivity and in-plane stress-strain response of the CNT mat and CNT-PANI electrodes were measured immediately after manufacture in the dry state. The tensile Young's modulus and yield strength depend upon the volume fraction of CNT bundles in a non-linear manner; the influence of PANI volume fraction is minor. Electrical conductivity scales linearly with measured CNT bundle volume fraction: it is largely unaffected by the value of PANI volume fraction. Both in-plane mechanical properties and the electrical conductivity vary with sample orientation, and possess a similar degree of anisotropy. The yield strength of PANI [70,71] is much below the shear yield strength of CNT bundles. Also, the electrical conductivity of PANI [70,71] is much below that measured for the mat and CNT-PANI electrodes. The minor influence of PANI upon the mechanical strength and electrical conductivity of the CNT mat and CNT-PANI electrodes implies that it does not penetrate the tightly-packed CNT bundles, and only coats the bundle surfaces. The X-ray diffraction spectra measured for the CNT-PANI electrodes and CNT mat absent PANI are alike, confirming that the internal bundle structure is unchanged by the addition of PANI.

The electrode capacitance was measured with cyclic voltammetry tests, and rises with increasing PANI mass fraction in a non-linear manner over the range of electrode compositions studied herein. Electrode capacitance is also dependent upon the scan rate of voltage during cyclic voltammetry, due to the time dependence of the Faradic reactions that take place upon charge and discharge.

Now consider the effect of pre-charging or cycling upon the electrochemical properties of the CNT-PANI composite electrodes. The mass of PANI within the electrodes is reduced by pre-charging or cycling. This decrease in mass upon oxidation is attributed to chemical reactions at the end of PANI molecular chains, which result in the formation of visible soluble degradation products [44-47] in solution around the immersed electrode. The soluble degradation products are removed by rinsing of pre-charged or cycled electrodes, and the remaining PANI upon the CNT bundles continues to act as active material. The relationship between the mass fraction of remaining PANI material within the electrodes and the electrode capacitance is identical to that measured for CNT-PANI electrodes immediately after manufacture. We conclude that the reduction in electrode capacitance upon pre-charging or cycling is directly attributable to the reduction in PANI mass.

Pre-charging and cycling also affect the mechanical properties of the CNT-PANI composites. Pre-cycling and charging lead to swelling of the electrodes in their through-thickness direction. Consequently, the CNT bundle volume fraction, modulus and strength are reduced from their initial values. A micromechanical model was constructed to predict the stress-strain response of the CNT-PANI electrodes as a function of CNT bundle volume fraction and PANI volume fraction; good agreement was obtained with experiment across the range of manufactured electrode compositions. The micromechanical model was also used to obtain predictions for the stress-strain response of pre-charged and pre-cycled electrodes, based upon their measured composition after pre-cycling and pre-charging. By accounting for the change in electrode composition due to pre-cycling and pre-charging the model was able to predict the observed reduction in electrode modulus and strength.

## Acknowledgements

Funding from the EPSRC project 'Advanced Nanotube Application and Manufacturing (ANAM) Initiative' under Grant No. EP/M015211/1, and from the ERC projects 'Multi-phase Lattice Materials' (MULTILAT) under Grant No. 669764, and 'Roll to Roll Manufacturing of Li Ion Battery Electrodes' (MIGHTY) UNDER Grant No. 866005 is gratefully acknowledged by all authors. The first author would thank for the financial support from the Cambridge CAPE Acorn Blue Sky Research Award (Grant No. NMZD/256). The work was also supported by the National Research Foundation of Korea (NRF) grant funded by the Korean government (No. 2020R1G1A1101146). All authors also thank Tortech Nano Fibers Ltd for providing the CNT mats used in experiment.

## References

[1]  C. Breitkopf, K. Swider-Lyons, Springer handbook of electrochemical energy, Springer, 2016.

[2]  L. Dai, D.W. Chang, J.B. Baek, W. Lu, Carbon nanomaterials for advanced energy conversion and storage, Small. 8 (2012) 1130–1166. https://doi.org/10.1002/smll.201101594.

[3]  M.F.L. De Volder, S.H. Tawfick, R.H. Baughman, A.J. Hart, Carbon Nanotubes: Present and Future Commercial Applications, Science (80-. ). 339 (2013) 535–539. https://doi.org/10.1126/science.1222453.

[4]  T.W. Chou, L. Gao, E.T. Thostenson, Z. Zhang, J.H. Byun, An assessment of the science and technology of carbon nanotube-based fibers and composites, Compos. Sci. Technol. 70 (2010) 1–19. https://doi.org/10.1016/j.compscitech.2009.10.004.

[5]  A. Pendashteh, E. Senokos, J. Palma, M. Anderson, J.J. Vilatela, R. Marcilla, Manganese dioxide decoration of macroscopic carbon nanotube fibers: From high-performance liquid-based to all-solid-state supercapacitors, J. Power Sources. 372 (2017) 64–73. https://doi.org/10.1016/J.JPOWSOUR.2017.10.068.

[6]   Y.-L. Li, I.A. Kinloch, A.H. Windle, Direct Spinning of Carbon Nanotube Fibers from Chemical Vapor Deposition Synthesis, Science (80-. ). 304 (2004) 276–278. https://doi.org/10.1126/science.1094982.

[7]   M. Kumar, Y. Ando, Chemical Vapor Deposition of Carbon Nanotubes: A Review on Growth Mechanism and Mass Production, J. Nanosci. Nanotechnol. 10 (2010) 3739–3758. https://doi.org/10.1166/jnn.2010.2939.

[8]   W. Tan, J.C. Stallard, F.R. Smail, A.M. Boies, N.A. Fleck, The mechanical and electrical properties of direct-spun carbon nanotube mat-epoxy composites, Carbon N. Y. 150 (2019) 489–504. https://doi.org/10.1016/j.carbon.2019.04.118.

[9]   J.C. Stallard, W. Tan, F.R. Smail, T.S. Gspann, A.M. Boies, N.A. Fleck, The mechanical and electrical properties of direct-spun carbon nanotube mats, Extrem. Mech. Lett. 21 (2018) 65–75. https://doi.org/10.1016/J.EML.2018.03.003.

[10]  A. Peigney, C. Laurent, E. Flahaut, R.R. Bacsa, A. Rousset, Specific surface area of carbon nanotubes and bundles of carbon nanotubes, Carbon N. Y. 39 (2001) 507–514. https://doi.org/10.1016/S0008-6223(00)00155-X.

[11]  C. Niu, E.K. Sichel, R. Hoch, D. Moy, H. Tennent, High power electrochemical capacitors based on carbon nanotube electrodes, Appl. Phys. Lett. 70 (1997) 1480–1482. https://doi.org/10.1063/1.118568.

[12]  K.H. An, W.S. Kim, Y.S. Park, J.M. Moon, D.J. Bae, S.C. Lim, Y.S. Lee, Y.H. Lee, Electrochemical properties of high-power supercapacitors using single-walled carbon nanotube electrodes, Adv. Funtional Mater. 11 (2001) 387–392. https://doi.org/10.1002/1616-3028(200110)11:5<387::AID-ADFM387>3.0.CO;2-G.

[13]  C.M. White, R. Banks, I. Hamerton, J.F. Watts, Characterisation of commercially CVD grown multi-walled carbon nanotubes for paint applications, Prog. Org. Coatings. 90 (2016) 44–53. https://doi.org/10.1016/j.porgcoat.2015.09.020.

[14]  E. Frackowiak, F. Béguin, Carbon materials for the electrochemical storage of energy in capacitors, Carbon N. Y. 39 (2001) 937–950. https://doi.org/10.1016/S0008-6223(00)00183-4.

[15]  C.V.V. Muralee Gopi, R. Vinodh, S. Sambasivam, I.M. Obaidat, H.J. Kim, Recent progress of advanced energy storage materials for flexible and wearable supercapacitor: From design and development to applications, J. Energy Storage. 27 (2020) 101035. https://doi.org/10.1016/j.est.2019.101035.

[16]  R. Perez-Gonzalez, Z. Peng, D. Camacho, A.I. Oliva, Q. Pei, A. Zakhidov, A. Encinas, J. Oliva, All solid state stretchable carbon nanotube based supercapacitors with controllable output voltage, J. Energy Storage. (2020). https://doi.org/10.1016/j.est.2020.101844.

[17]  J.W. Kim, E.J. Siochi, J. Carpena-Núñez, K.E. Wise, J.W. Connell, Y. Lin, R.A. Wincheski, Polyaniline/carbon nanotube sheet nanocomposites: Fabrication and characterization, ACS Appl. Mater. Interfaces. 5 (2013) 8597–8606. https://doi.org/10.1021/am402077d.

[18]  E. Senokos, Y. Ou, J.J. Torres, F. Sket, C. González, R. Marcilla, J.J. Vilatela, Energy storage in structural composites by introducing CNT fiber/polymer electrolyte interleaves, Sci. Rep. 8 (2018) 3407–3417. https://doi.org/10.1038/s41598-018-21829-5.

[19]  N. Shirshova, H. Qian, M.S.P. Shaffer, J.H.G. Steinke, E.S. Greenhalgh, P.T. Curtis, A. Kucernak, A. Bismarck, Structural composite supercapacitors, Compos. Part A Appl. Sci. Manuf. 46 (2013) 96–107. https://doi.org/10.1016/j.compositesa.2012.10.007.

[20]  V. Gupta, N. Miura, Polyaniline/single-wall carbon nanotube (PANI/SWCNT) composites for high performance supercapacitors, Electrochim. Acta. 52 (2006) 1721–1726. https://doi.org/10.1016/j.electacta.2006.01.074.

[21]  Y. Zhou, Z.Y. Qin, L. Li, Y. Zhang, Y.L. Wei, L.F. Wang, M.F. Zhu, Polyaniline/multi-walled carbon nanotube composites with core-shell structures as supercapacitor electrode

materials, Electrochim. Acta. 55 (2010) 3904–3908. https://doi.org/10.1016/j.electacta.2010.02.022.

[22] R. Oraon, A. De Adhikari, S.K. Tiwari, G.C. Nayak, Enhanced Specific Capacitance of Self-Assembled Three-Dimensional Carbon Nanotube/Layered Silicate/Polyaniline Hybrid Sandwiched Nanocomposite for Supercapacitor Applications, ACS Sustain. Chem. Eng. 4 (2016) 1392–1403. https://doi.org/10.1021/acssuschemeng.5b01389.

[23] Z. Niu, P. Luan, Q. Shao, H. Dong, J. Li, J. Chen, D. Zhao, L. Cai, W. Zhou, X. Chen, S. Xie, A '"skeleton/skin"' strategy for preparing ultrathin free-standing single-walled carbon nanotube/polyaniline films for high performance supercapacitor electrodes, Energy Environ. Sci. This J. Is [a] R. Soc. Chem. Energy Environ. Sci. 8726 (2012) 8726–8733. https://doi.org/10.1039/c2ee22042c.

[24] J. Zhang, L.-B. Kong, B. Wang, Y.-C. Luo, L. Kang, In-situ electrochemical polymerization of multi-walled carbon nanotube/polyaniline composite films for electrochemical supercapacitors, Synth. Met. 159 (2009) 260–266. https://doi.org/10.1016/j.synthmet.2008.09.018.

[25] J. Yu, W. Lu, S. Pei, K. Gong, L. Wang, L. Meng, Y. Huang, J.P. Smith, K.S. Booksh, Q. Li, J.-H. Byun, Y. Oh, Y. Yan, T.-W. Chou, Omnidirectionally Stretchable High-Performance Supercapacitor Based on Isotropic Buckled Carbon Nanotube Films, (2016). https://doi.org/10.1021/acsnano.6b00752.

[26] C. Meng, C. Liu, S. Fan, Flexible carbon nanotube/polyaniline paper-like films and their enhanced electrochemical properties, Electrochem. Commun. 11 (2009) 186–189. https://doi.org/10.1016/j.elecom.2008.11.005.

[27] P. Liu, J. Yan, Z. Guang, Y. Huang, X. Li, W. Huang, Recent advancements of polyaniline-based nanocomposites for supercapacitors, J. Power Sources. 424 (2019) 108–130. https://doi.org/10.1016/j.jpowsour.2019.03.094.

[28] D.C. Nguyen, D.T. Tran, T.L.L. Doan, D.H. Kim, N.H. Kim, J.H. Lee, Rational Design of Core@shell Structured $CoS_x$@$Cu_2MoS_4$ Hybridized $MoS_2$/N,S‐Codoped Graphene as Advanced Electrocatalyst for Water Splitting and Zn‐Air Battery, Adv. Energy Mater. 10 (2020) 1903289. https://doi.org/10.1002/aenm.201903289.

[29] J. Du, L. Liu, Y. Yu, Y. Zhang, H. Lv, A. Chen, N-doped ordered mesoporous carbon spheres derived by confined pyrolysis for high supercapacitor performance, J. Mater. Sci. Technol. 35 (2019) 2178–2186. https://doi.org/10.1016/j.jmst.2019.05.029.

[30] M. Chen, Y. Zhang, Y. Liu, Q. Wang, J. Zheng, C. Meng, Three-Dimensional Network of Vanadium Oxyhydroxide Nanowires Hybridize with Carbonaceous Materials with Enhanced Electrochemical Performance for Supercapacitor, ACS Appl. Energy Mater. 1 (2018) 5527–5538. https://doi.org/10.1021/acsaem.8b01109.

[31] T. Hu, Y. Liu, Y. Zhang, M. Chen, J. Zheng, J. Tang, C. Meng, 3D hierarchical porous $V_3O_7 \cdot H_2O$ nanobelts/CNT/reduced graphene oxide integrated composite with synergistic effect for supercapacitors with high capacitance and long cycling life, J. Colloid Interface Sci. 531 (2018) 382–393. https://doi.org/10.1016/j.jcis.2018.07.060.

[32] H. Li, J. Wang, Q. Chu, Z. Wang, F. Zhang, S. Wang, Theoretical and experimental specific capacitance of polyaniline in sulfuric acid, J. Power Sources. 190 (2009) 578–586. https://doi.org/10.1016/j.jpowsour.2009.01.052.

[33] M.A. Bavio, G.G. Acosta, T. Kessler, Synthesis and characterization of polyaniline and polyaniline - Carbon nanotubes nanostructures for electrochemical supercapacitors, J. Power Sources. 245 (2014) 475–481. https://doi.org/10.1016/j.jpowsour.2013.06.119.

[34] S. Quillard, G. Louarn, S. Lefrant, A.G. Macdiarmid, Vibrational analysis of polyaniline: A comparative study of leucoemeraldine, emeraldine, and pernigraniline bases, Phys. Rev. 50 (1994) 12496–12508.

[35] S.B. Yoon, E.H. Yoon, K.B. Kim, Electrochemical properties of leucoemeraldine, emeraldine, and pernigraniline forms of polyaniline/multi-wall carbon nanotube nanocomposites for supercapacitor applications, J. Power Sources. 196 (2011) 10791–10797. https://doi.org/10.1016/j.jpowsour.2011.08.107.

[36] A. Eftekhari, L. Li, Y. Yang, Polyaniline supercapacitors, J. Power Sources. 347 (2017) 86–107. https://doi.org/10.1016/j.jpowsour.2017.02.054.

[37] E.N. Konyushenko, J. Stejskal, M. Trchová, J. Hradil, J. Kovářová, J. Prokeš, M. Cieslar, J.Y. Hwang, K.H. Chen, I. Sapurina, Multi-wall carbon nanotubes coated with polyaniline, Polymer (Guildf). 47 (2006) 5715–5723. https://doi.org/10.1016/j.polymer.2006.05.059.

[38] V. Mottaghitalab, G.M. Spinks, G.G. Wallace, The influence of carbon nanotubes on mechanical and electrical properties of polyaniline fibers, Synth. Met. 152 (2005) 77–80. https://doi.org/10.1016/j.synthmet.2005.07.154.

[39] H. Zhang, G. Cao, W. Wang, K. Yuan, B. Xu, W. Zhang, J. Cheng, Y. Yang, Influence of microstructure on the capacitive performance of polyaniline/carbon nanotube array composite electrodes, Electrochim. Acta. 54 (2009) 1153–1159. https://doi.org/10.1016/j.electacta.2008.09.004.

[40] S. Jiao, T. Li, Y. Zhang, C. Xiong, T. Zhao, M. Khan, A three-dimensional vertically aligned carbon nanotube/polyaniline composite as a supercapacitor electrode, (2016). https://doi.org/10.1039/c6ra17674g.

[41] D. Zhang, M. Miao, H. Niu, Z. Wei, Core-Spun Carbon Nanotube Yarn Supercapacitors for Wearable Electronic Textiles, 16 (2019) 38. https://doi.org/10.1021/nn5001386.

[42] B. Wang, X. Fang, H. Sun, S. He, J. Ren, Y. Zhang, H. Peng, Fabricating Continuous Supercapacitor Fibers with High Performances by Integrating All Building Materials and Steps into One Process, Adv. Mater. 27 (2015) 7854–7860. https://doi.org/10.1002/adma.201503441.

[43] C. Meng, C. Liu, L. Chen, C. Hu, S. Fan, Highly flexible and all-solid-state paperlike polymer supercapacitors, Nano Lett. 10 (2010) 4025–4031. https://doi.org/10.1021/nl1019672.

[44] D.E. Tsentalovich, R.J. Headrick, F. Mirri, J. Hao, N. Behabtu, C.C. Young, M. Pasquali, Influence of Carbon Nanotube Characteristics on Macroscopic Fiber Properties, (n.d.). https://doi.org/10.1021/acsami.7b10968.

[45] N. Behabtu, M.J. Green, M. Pasquali, Carbon nanotube-based neat fibers, Nano Today. 3 (2008) 24–34. https://doi.org/10.1016/S1748-0132(08)70062-8.

[46] M. Endo, T. Hayashi, Y.A. Kim, Large-scale production of carbon nanotubes and their applications, in: Pure Appl. Chem., De Gruyter, 2006: pp. 1703–1713. https://doi.org/10.1351/pac200678091703.

[47] M. Fathi, M. Saghafi, F. Mahboubi, S. Mohajerzadeh, Synthesis and electrochemical investigation of polyaniline/unzipped carbon nanotube composites as electrode material in supercapacitors, Synth. Met. 198 (2014) 345–356. https://doi.org/10.1016/j.synthmet.2014.10.033.

[48] J. Yu, W. Lu, S. Pei, K. Gong, L. Wang, L. Meng, Y. Huang, J.P. Smith, K.S. Booksh, Q. Li, J.H. Byun, Y. Oh, Y. Yan, T.W. Chou, Omnidirectionally Stretchable High-Performance Supercapacitor Based on Isotropic Buckled Carbon Nanotube Films, ACS Nano. 10 (2016) 5204–5211. https://doi.org/10.1021/acsnano.6b00752.

[49] A. González, E. Goikolea, J.A. Barrena, R. Mysyk, Review on supercapacitors: Technologies and materials, Renew. Sustain. Energy Rev. 58 (2016) 1189–1206. https://doi.org/10.1016/j.rser.2015.12.249.

[50] P. Sharma, T.S. Bhatti, A review on electrochemical double-layer capacitors, Energy Convers. Manag. 51 (2010) 2901–2912. https://doi.org/10.1016/j.enconman.2010.06.031.

[51] P.Y. Chen, N.M. Dorval Courchesne, M.N. Hyder, J. Qi, A.M. Belcher, P.T. Hammond, Carbon nanotube-polyaniline core-shell nanostructured hydrogel for electrochemical energy storage, RSC Adv. 5 (2015) 37970–37977. https://doi.org/10.1039/c5ra02944a.

[52] J. Liu, M. Zhou, L.Z. Fan, P. Li, X. Qu, Porous polyaniline exhibits highly enhanced electrochemical capacitance performance, Electrochim. Acta. 55 (2010) 5819–5822. https://doi.org/10.1016/j.electacta.2010.05.030.

[53] M. Yoshio, R.J. Brodd, A. Kozawa, Lithium-ion batteries: Science and technologies, 2009. https://doi.org/10.1007/978-0-387-34445-4.

[54] M. Kalaji, L. Nyholm, L.M. Peter, A microelectrode study of the influence of pH and solution composition on the electrochemical behaviour of polyaniline films, J. Electroanal. Chem. 313 (1991) 271–289. https://doi.org/10.1016/0022-0728(91)85185-R.

[55] B.E. Conway, Electrochemical Supercapacitors : Scientific Fundamentals and Technological Applications, Springer, 2013.

[56] M.A. Azam, N.S.A. Manaf, E. Talib, M.S.A. Bistamam, Aligned carbon nanotube from catalytic chemical vapor deposition technique for energy storage device: A review, Ionics (Kiel). 19 (2013) 1455–1476. https://doi.org/10.1007/s11581-013-0979-x.

[57] L.D. Arsov, W. Plieth, G. Koßmehl, Electrochemical and Raman spectroscopic study of polyaniline; Influence of the potential on the degradation of polyaniline, J. Solid State Electrochem. 2 (1998) 355–361. https://doi.org/10.1007/s100080050112.

[58] C. Hoecker, F. Smail, M. Bajada, M. Pick, A. Boies, Catalyst nanoparticle growth dynamics and their influence on product morphology in a CVD process for continuous carbon nanotube synthesis, Carbon N. Y. 96 (2016) 116–124. https://doi.org/10.1016/j.carbon.2015.09.050.

[59] B. Alemán, V. Reguero, B. Mas, J.J. Vilatela, Strong Carbon Nanotube Fibers by Drawing Inspiration from Polymer Fiber Spinning, ACS Nano. 9 (2015) 7392–7398. https://doi.org/10.1021/acsnano.5b02408.

[60] T. Gspann, F. Smail, A. Windle, Spinning of carbon nanotube fibres using the floating catalyst high temperature route: purity issues and the critical role of sulphur: Supporting Online Material 1, Faraday Discuss. 173 (2014) 2–7. https://doi.org/10.1039/C4FD00066H.

[61] J. Stejskal, R.G. Gilbert, Polyaniline. Preparation of a conducting polymer (IUPAC technical report), Pure Appl. Chem. 74 (2002) 857–867. https://doi.org/10.1351/pac200274050857.

[62] X. Zhao, Y. Ando, Raman spectra and X-ray diffraction patterns of carbon nanotubes prepared by hydrogen arc discharge, Japanese J. Appl. Physics, Part 1 Regul. Pap. Short Notes Rev. Pap. 37 (1998) 4846–4849. https://doi.org/10.1143/jjap.37.4846.

[63] A.G. MacDiarmid, A.J. Epstein, Polyanilines: A novel class of conducting polymers, Faraday Discuss. Chem. Soc. 88 (1989) 317–332. https://doi.org/10.1039/DC9898800317.

[64] A. Andreatta, Y. Cao, J.C. Chiang, A.J. Heeger, P. Smith, Electrically-conductive fibers of polyaniline spun from solutions in concentrated sulfuric acid, Synth. Met. 26 (1988) 383–389. https://doi.org/10.1016/0379-6779(88)90233-0.

[65] G. Wu, P. Tan, D. Wang, Z. Li, L. Peng, Y. Hu, C. Wang, W. Zhu, S. Chen, W. Chen, High-performance Supercapacitors Based on Electrochemical-induced Vertical-aligned Carbon Nanotubes and Polyaniline Nanocomposite Electrodes, Sci. Rep. 7 (2017). https://doi.org/10.1038/srep43676.

[66] J. Xing, M. Liao, C. Zhang, M. Yin, D. Li, Y. Song, The effect of anions on the electrochemical properties of polyaniline for supercapacitors, Phys. Chem. Chem. Phys. 19 (2017) 14030–14041. https://doi.org/10.1039/c7cp02016c.

[67] S.M. Park, Electrochemistry of Conductive Polymers III. Some Physical and Electrochemical Properties Observed from Electrochemically Grown Polyaniline, J. Electrochem. Soc. 135 (1988) 2491–2496. https://doi.org/10.1149/1.2095364.

[68] E.M. Geniès, M. Lapkowski, J.F. Penneau, Cyclic voltammetry of polyaniline: interpretation of the middle peak, J. Electroanal. Chem. 249 (1988) 97–107. https://doi.org/10.1016/0022-0728(88)80351-6.

[69] L. Lizarraga, E.M. Andrade, F.V. Molina, Swelling and volume changes of polyaniline upon redox switching, J. Electroanal. Chem. 561 (2004) 127–135. https://doi.org/10.1016/j.jelechem.2003.07.026.

[70] S.J. Pomfret, P.N. Adams, N.P. Comfort, A.P. Monkman, Electrical and mechanical properties of polyaniline fibres produced by a one-step wet spinning process, Polymer (Guildf). 41 (2000) 2265–2269. https://doi.org/10.1016/S0032-3861(99)00365-1.

[71] H. Valentová, J. Stejskal, Mechanical properties of polyaniline, Synth. Met. 160 (2010) 832–834. https://doi.org/10.1016/j.synthmet.2010.01.007.

[72] J.-P. Salvetat, G. Briggs, J.-M. Bonard, R. Bacsa, A. Kulik, T. Stöckli, N. Burnham, L. Forró, Elastic and Shear Moduli of Single-Walled Carbon Nanotube Ropes, Phys. Rev. Lett. 82 (1999) 944–947. https://doi.org/10.1103/PhysRevLett.82.944.

[73] A. Kis, G. Csányi, J.-P. Salvetat, T.-N. Lee, E. Couteau, A.J. Kulik, W. Benoit, J. Brugger, L. Forró, Reinforcement of single-walled carbon nanotube bundles by intertube bridging, Nat. Mater. 3 (2004) 153–157. https://doi.org/10.1038/nmat1076.

[74] B. Peng, M. Locascio, P. Zapol, S. Li, S.L. Mielke, G.C. Schatz, H.D. Espinosa, Measurements of near-ultimate strength for multiwalled carbon nanotubes and irradiation-induced crosslinking improvements, Nat. Nanotechnol. 3 (2008) 626–631. https://doi.org/10.1038/nnano.2008.211.

[75] R. Hill, A Theory of the Yielding and Plastic Flow of Anisotropic Metals, Proc. R. Soc. A Math. Phys. Eng. Sci. 193 (1948) 281–297. https://doi.org/10.1098/rspa.1948.0045.

[76] M. Naraghi, T. Filleter, A. Moravsky, M. Locascio, R.O. Loutfy, H.D. Espinosa, A Multiscale Study of High Performance Double-Walled Nanotube−Polymer Fibers, ACS Nano. 4 (2010) 6463–6476. https://doi.org/10.1021/nn101404u.

# Tables

Table 1: Effect of deposition time upon CNT mat and CNT-PANI composite composition

| Sample label | Deposition time (min) | Density $\rho$ (kg m$^{-3}$) | PANI volume fraction $f_P$ | CNT bundle volume fraction $f_B$ | Void volume fraction $f_a$ |
|---|---|---|---|---|---|
| CNT mat | 0 | 292 | 0 | 0.187 | 0.813 |
| CNT-PANI-T5 | 5 | 335 | 0.015 | 0.203 | 0.778 |
| CNT-PANI-T10 | 10 | 391 | 0.028 | 0.228 | 0.745 |
| CNT-PANI-T20 | 20 | 500 | 0.087 | 0.251 | 0.662 |
| CNT-PANI-T30 | 30 | 606 | 0.140 | 0.277 | 0.583 |
| CNT-PANI-T60 | 60 | 823 | 0.294 | 0.293 | 0.413 |

Table 2: Composition of CNT-PANI-T20 composites after pre-charging or cycling

| Sample label | Density $\rho$ (kg m$^{-3}$) | PANI volume fraction $f_P$ | CNT bundle volume fraction $f_B$ | Void volume fraction $f_a$ |
|---|---|---|---|---|
| 0.43 V (OCV) | 500 | 0.087 | 0.251 | 0.662 |
| 0.43 V (1000 cycles) | 432 | 0.069 | 0.222 | 0.708 |
| -0.2 V | 412 | 0.060 | 0.216 | 0.724 |
| 0.1 V | 485 | 0.081 | 0.246 | 0.673 |
| 0.3 V | 478 | 0.082 | 0.241 | 0.677 |
| 0.6 V | 466 | 0.074 | 0.240 | 0.686 |
| 1 V | 318 | 0.015 | 0.192 | 0.793 |

Table 3: Material parameters adopted in finite element computation

| Phase | Material parameters |
|---|---|
| CNT bundle | $E_{11}^B = 680$ GPa, $E_{22}^B = E_{33}^B = 50$ GPa $G_{12}^B = G_{23}^B = G_{13}^B = 9.5$ GPa, $\nu_{12} = \nu_{13} = 0.3$ $\sigma_{11}^B = 3700$ MPa $\sigma_{22}^B = \sigma_{33}^B = \tau_y^B = 480$ MPa $\tau_{12}^B = \tau_{23}^B = \tau_{13}^B = \tau_y^B$ |
| PANI-(M) | $E^P = 1.4$ GPa, $\nu_P = 0.3$, $\sigma_y^P = 83$ MPa [70,71] |

# Figure captions

Figure 1: (a) Typical configuration of a supercapacitor. (b) Chemical structures of the six different polyaniline forms.

Figure 2: (a) Classes of CNT-PANI composites. (b) The specific power and specific energy of CNT-PANI composites, capacitors [49], electrostatic double-layer supercapacitors (EDLC) [21,47,50] and lithium-ion batteries [53]. (c) The specific capacitance of CNT, porous PANI and CNT-PANI composites is plotted versus the volume fraction of PANI. The sources of data are [21,47] for CNT EDLCs, from for PANI [24,26,32,47,51,52], from [20–22,24,37,47] for the powder-form CNT-PANI composites, from [40,41] for the forest-spun CNT-PANI composites, from [17,23,43,48] for the direct-spun CNT-PANI composites and from the present study.

Figure 3: (a) Three-electrode setup for the CNT-PANI electrode manufacture and electrochemical characterisation. (b) The PANI mass fraction $m_P$ in CNT-PANI composite electrodes is plotted against the electrodeposition time. (c) Plan-view of the CNT mat bundle microstructure absent PANI. Plan views of the microstructure of CNT-PANI composite electrodes with increasing PANI content are shown in (d) to (h) as follows: (d) CNT-PANI-T5, (e) CNT-PANI-T10, (f) CNT-PANI-T20, (g) CNT-PANI-T30, (h) CNT-PANI-T60. Images (c) to (h) were obtained from scanning electron microscopy.

Figure 4: (a) Electrical conductivity measured CNT-PANI electrodes and CNT mat plotted against CNT volume fraction. (b) Voltage waveform applied during cyclic voltammetry. (c) Measured current $I$ and (d) measured charge $Q$ against voltage $\Phi$ during cyclic voltammetry. (e) Electrical capacitance versus PANI mass and (f) conversion efficiency versus PANI volume fraction measured in cyclic voltammetry for different scan rates.

Figure 5: Stress versus strain response of (a) direct-spun mat and CNT-PANI composite electrodes measured in line with the principal material orientation, and (b) of the CNT mat and CNT-PANI-T60 composites at angles of 0° and 90° to the principal material direction. (c) Young's modulus $E$ and (d) yield strength $\sigma_{YS}$ for CNT-PANI electrodes plotted against their density $\rho$. All data in (a) to (d) are measured after manufacture in the dry state.

Figure 6: (a) Voltage $\Phi$ applied during electrode pre-charging. (b) Current $I$ and (c) charge $Q$ plotted against applied voltage $\Phi$ during cyclic voltammetry testing of pre-charged CNT-PANI-T20 electrodes and electrodes subjected to 1000 charge/discharge cycles. (d) The capacitance of electrodes measured after manufacture and after the application of pre-charging or cycling, plotted against PANI mass fraction. Microstructure of CNT-PANI-T20 electrodes pre-charged to voltages of (e) 1 V and (f) -0.2 V. Images (e) and (f) obtained by scanning electron microscopy.

Figure 7: (a) Stress-strain response of charged and pre-charged CNT-PANI electrodes, (b) stress-strain response of CNT mat and CNT-PANI electrodes measured after manufacture and after the application of 1000 charge-discharge cycles.

Figure 8: (a) The planar honeycomb unit cell idealisation of CNT-PANI composite electrode microstructure. (b) Unit cell boundary conditions used for analysis with finite element

simulation. The measured and predicted uniaxial stress-strain responses of (c) as-manufactured CNT-PANI electrodes at 0° and at 90° to the principal direction, and (d) in the principal direction after the application of 1000 charge-discharge cycles or pre-charging to $\Phi_{PRE} = 1$ V and $\Phi_{PRE} = -0.2$ V. The (e) Young's modulus and (f) yield strength for CNT mat ($f_P = 0$) and for CNT-PANI composite electrodes as measured in experiment and as predicted with finite element simulation. Finite element predictions are included for PANI volume fractions $f_P = 0$ and $f_P = 0.30$.

# Figures

(a)

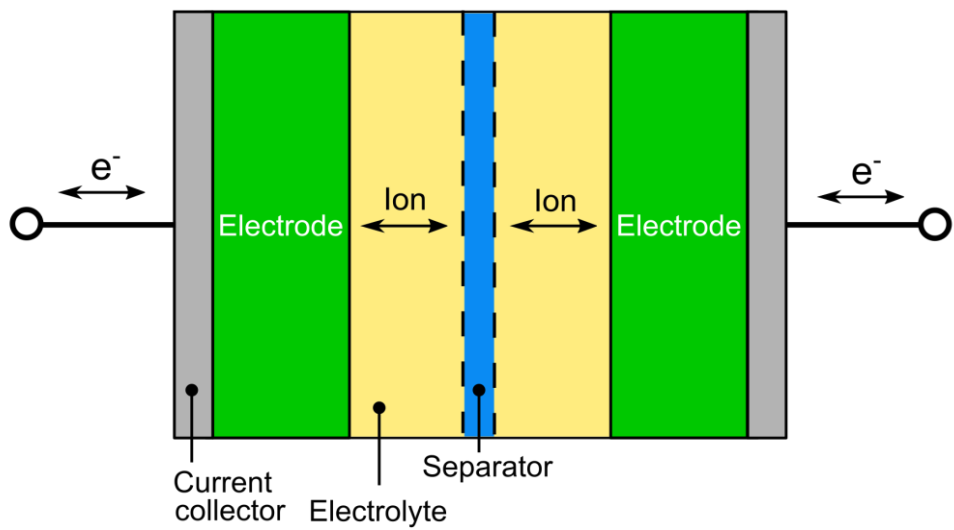

(b)

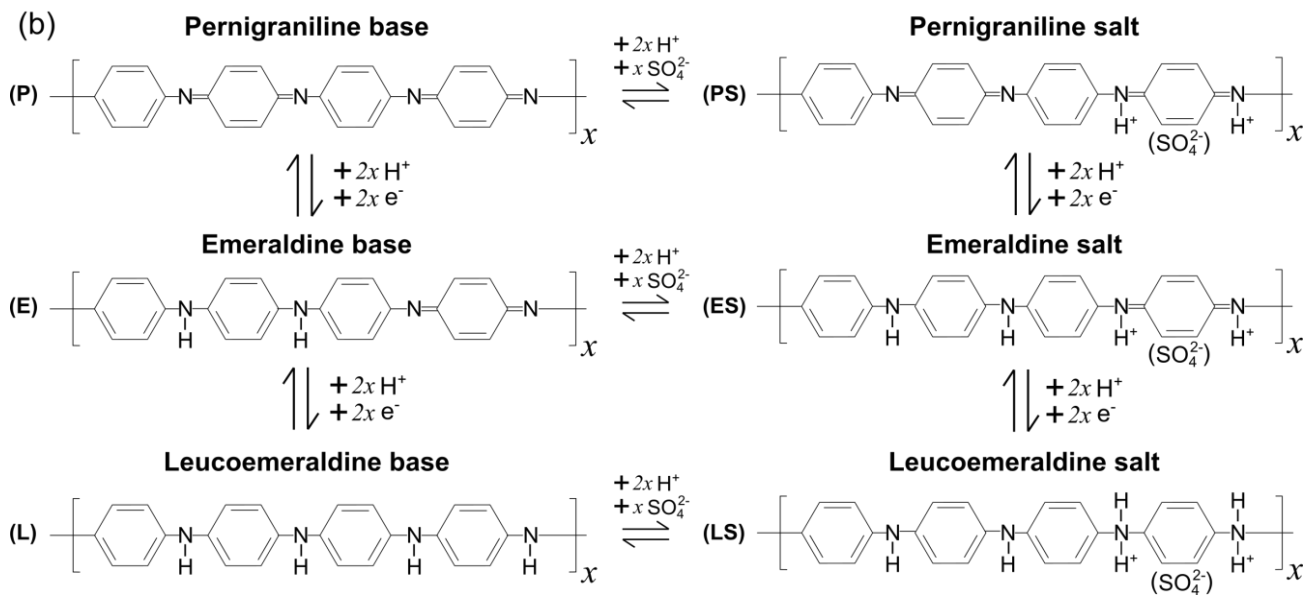

Figure 1: (a) Typical configuration of a supercapacitor. (b) Chemical structures of the six different polyaniline forms.

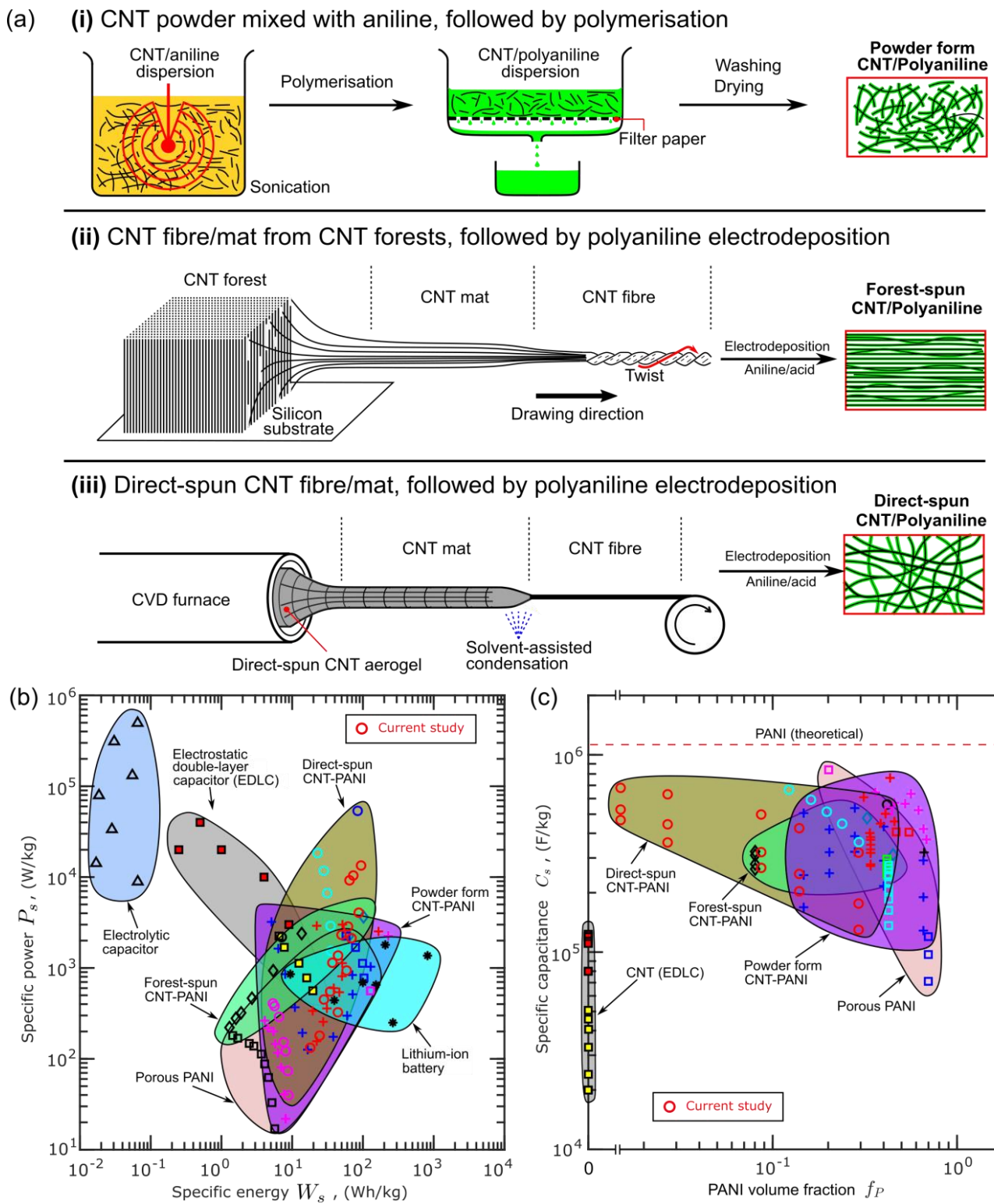

Figure 2: (a) Classes of CNT-PANI composites. (b) The specific power and specific energy of CNT-PANI composites, capacitors [49], electrostatic double-layer supercapacitors (EDLC) [21,47,50] and lithium-ion batteries [53]. (c) The specific capacitance of CNT, porous PANI and CNT-PANI composites is plotted versus the volume fraction of PANI. The sources of data are [21,47] for CNT EDLCs, from for PANI [24,26,32,47,51,52], from [20–22,24,37,47] for the powder-form CNT-PANI composites, from [40,41] for the forest-spun CNT-PANI composites, from [17,23,43,48] for the direct-spun CNT-PANI composites and from the present study.

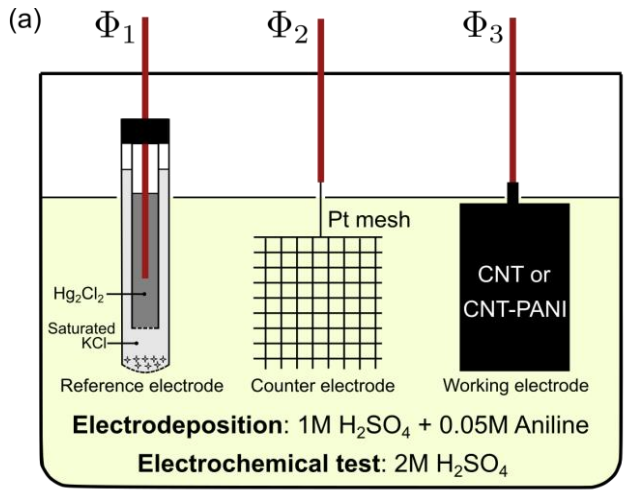
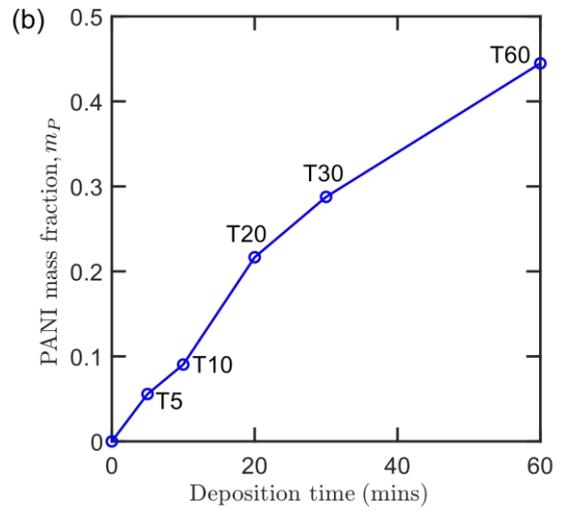
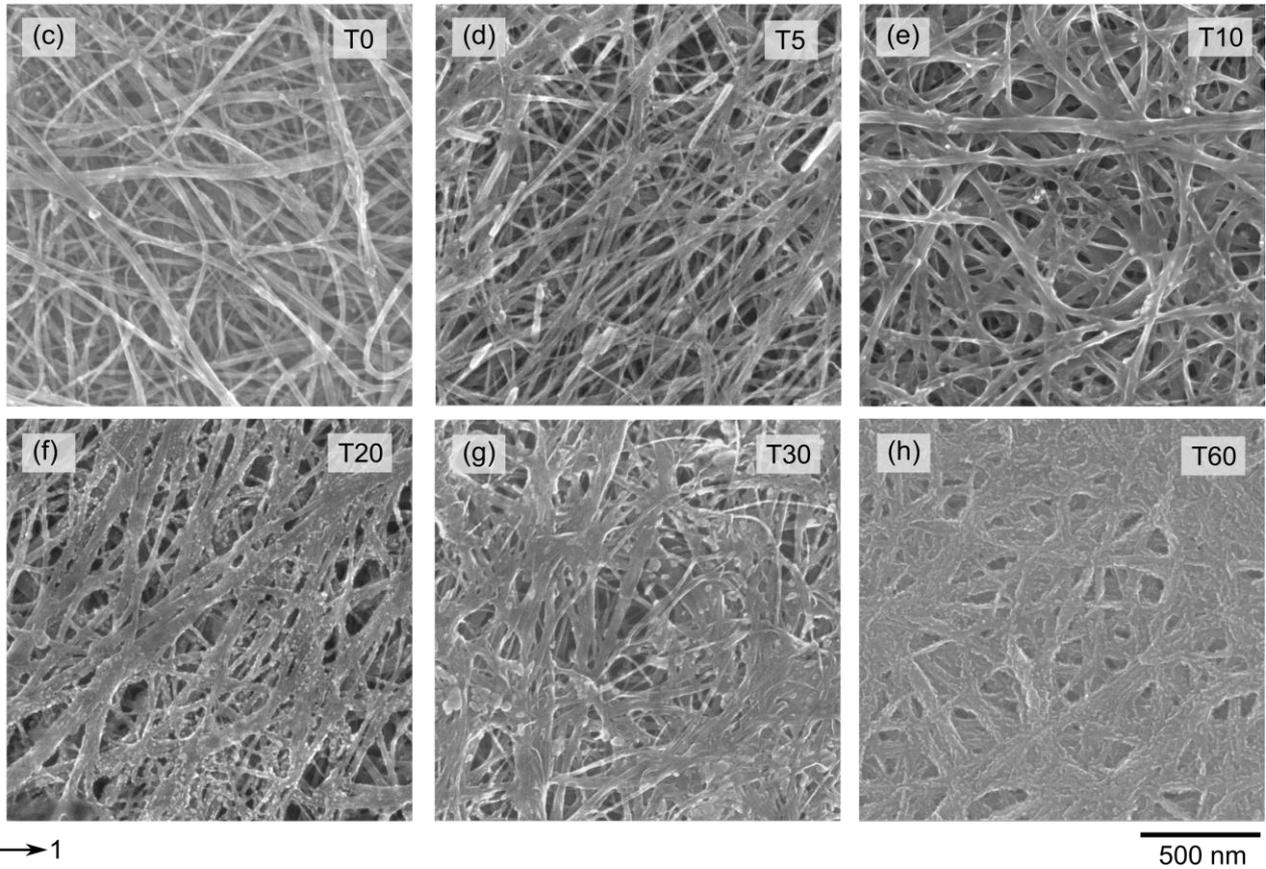

Figure 3: (a) Three-electrode setup for the CNT-PANI electrode manufacture and electrochemical characterisation. (b) The PANI mass fraction $m_P$ in CNT-PANI composite electrodes is plotted against the electrodeposition time. (c) Plan-view of the CNT mat bundle microstructure absent PANI. Plan views of the microstructure of CNT-PANI composite electrodes with increasing PANI content are shown in (d) to (h) as follows: (d) CNT-PANI-T5, (e) CNT-PANI-T10, (f) CNT-PANI-T20, (g) CNT-PANI-T30, (h) CNT-PANI-T60. Images (c) to (h) were obtained from scanning electron microscopy.

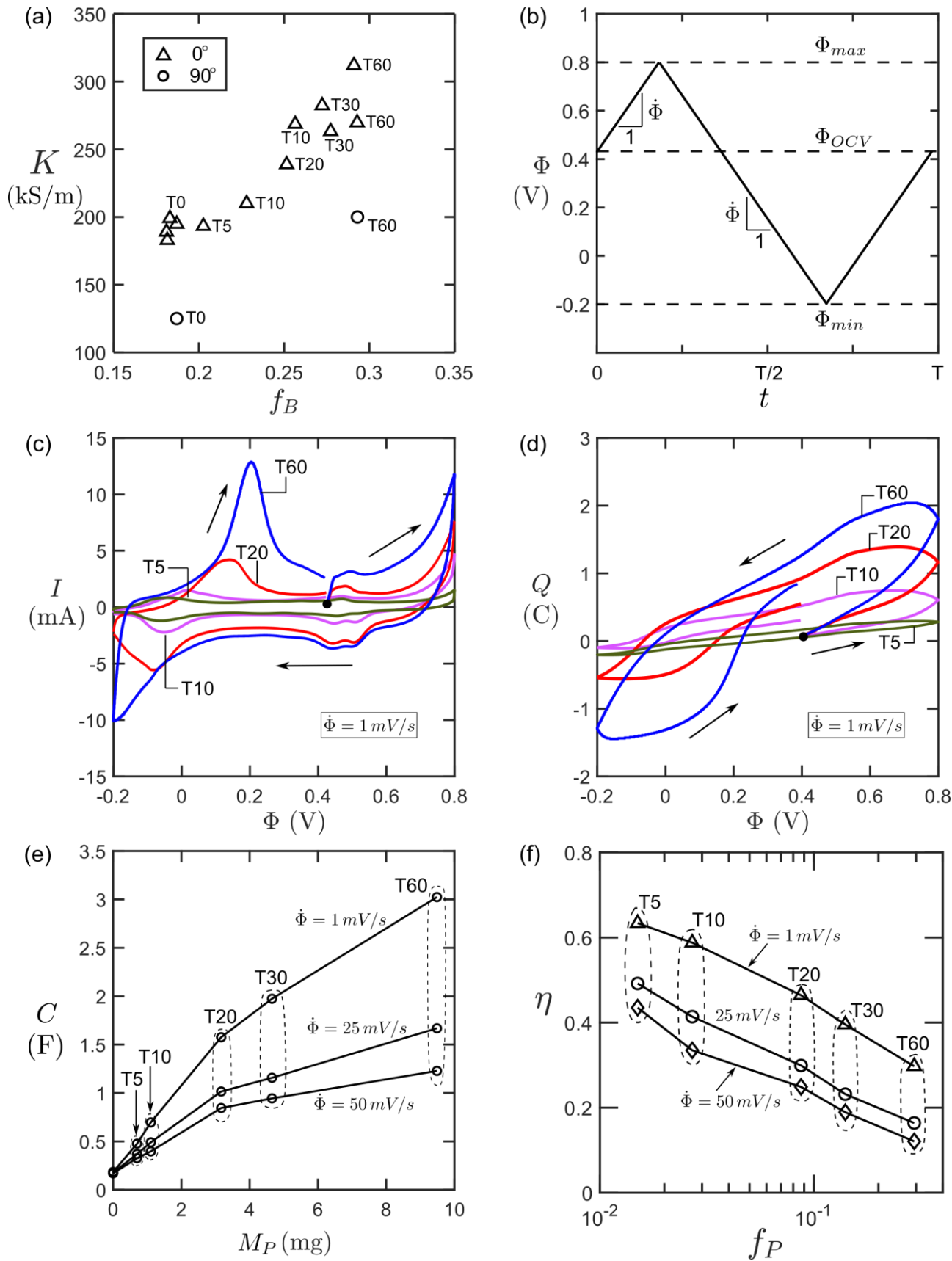

Figure 4: (a) Electrical conductivity measured CNT-PANI electrodes and CNT mat plotted against CNT volume fraction. (b) Voltage waveform applied during cyclic voltammetry. (c) Measured current $I$ and (d) measured charge $Q$ against voltage $\Phi$ during cyclic voltammetry. (e) Electrical capacitance versus PANI mass and (f) conversion efficiency versus PANI volume fraction measured in cyclic voltammetry for different scan rates.

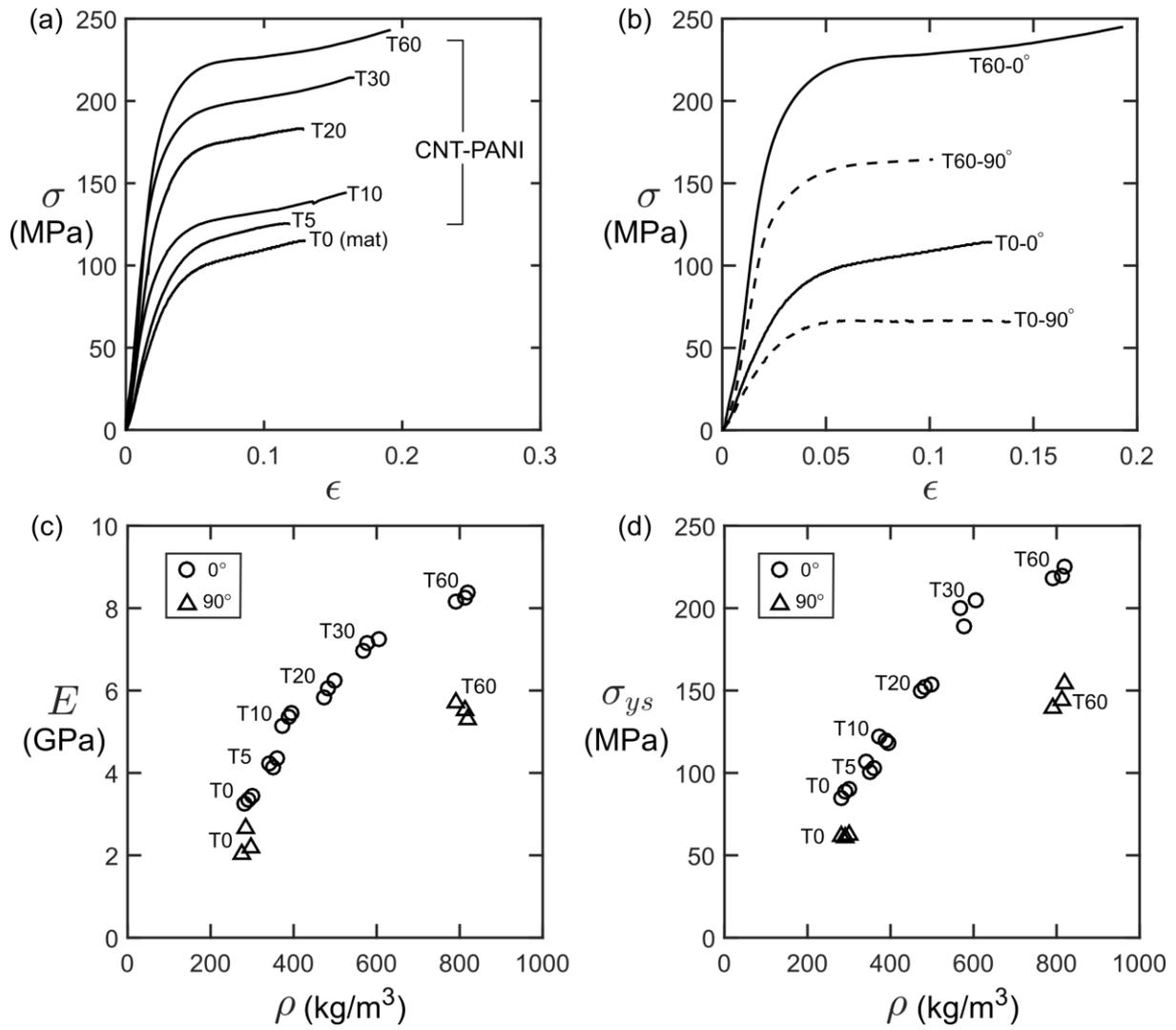

Figure 5: Stress versus strain response of (a) direct-spun mat and CNT-PANI composite electrodes measured in line with the principal material orientation, and (b) of the CNT mat and CNT-PANI-T60 composites at angles of 0° and 90° to the principal material direction. (c) Young's modulus $E$ and (d) yield strength $\sigma_{YS}$ for CNT-PANI electrodes plotted against their density $\rho$. All data in (a) to (d) are measured after manufacture in the dry state.

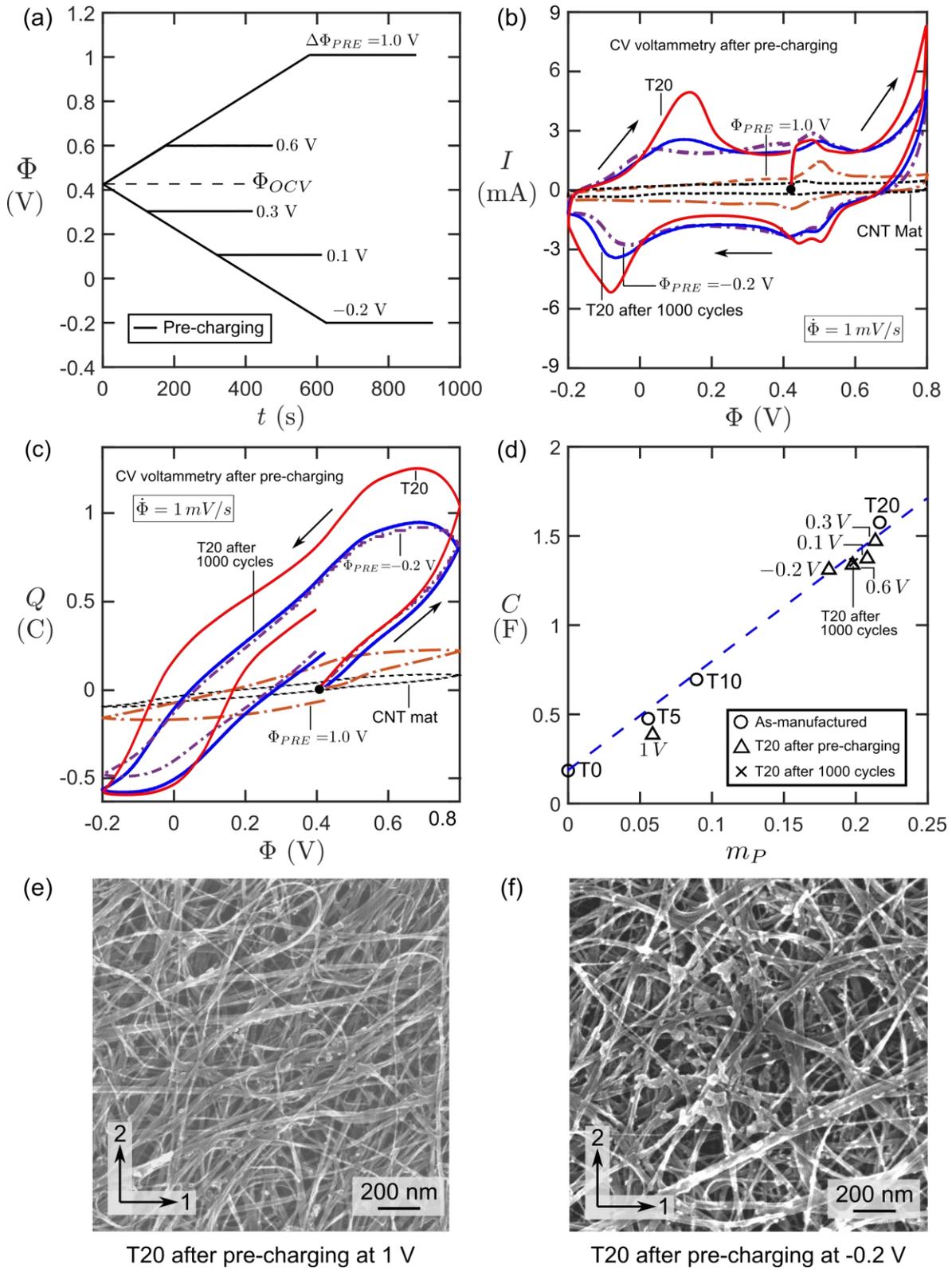

Figure 6: (a) Voltage $\Phi$ applied during electrode pre-charging. (b) Current $I$ and (c) charge $Q$ plotted against applied voltage $\Phi$ during cyclic voltammetry testing of pre-charged CNT-PANI-T20 electrodes and electrodes subjected to 1000 charge/discharge cycles. (d) The capacitance of electrodes measured after manufacture and after the application of pre-charging or cycling, plotted against PANI mass fraction. Microstructure of CNT-PANI-T20 electrodes pre-charged to voltages of (e) 1 V and (f) -0.2 V. Images (e) and (f) obtained by scanning electron microscopy.

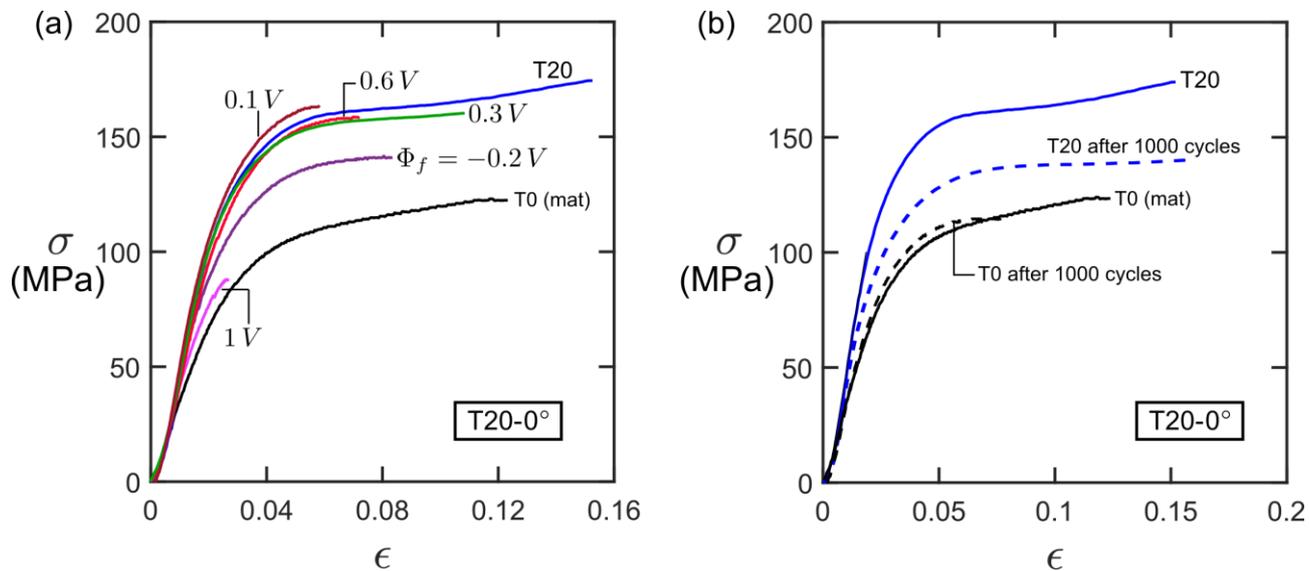

Figure 7: (a) Stress-strain response of charged and pre-charged CNT-PANI electrodes, (b) stress-strain response of CNT mat and CNT-PANI electrodes measured after manufacture and after the application of 1000 charge-discharge cycles.

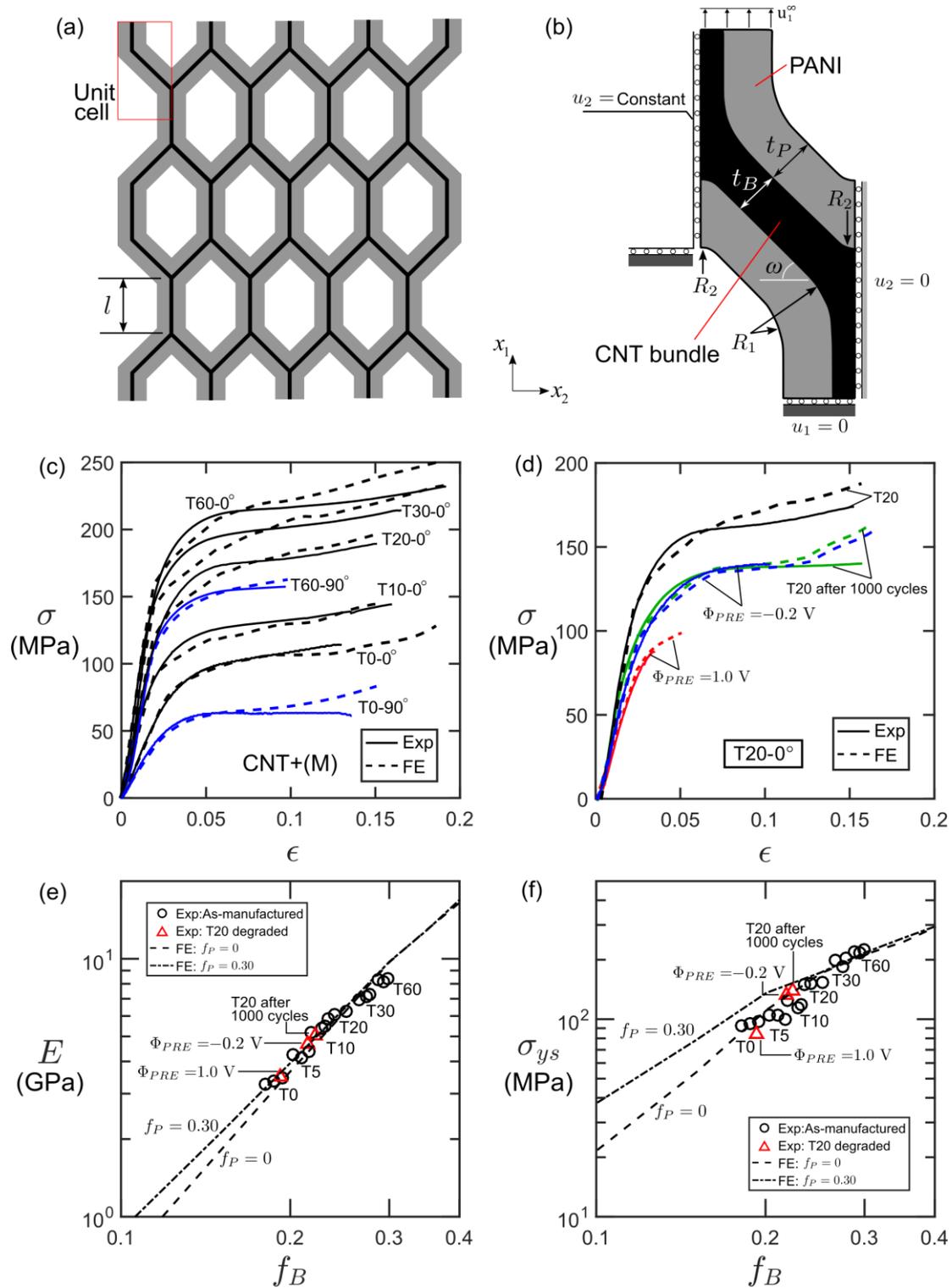

Figure 8: (a) The planar honeycomb unit cell idealisation of CNT-PANI composite electrode microstructure. (b) Unit cell boundary conditions used for analysis with finite element simulation. The measured and predicted uniaxial stress-strain responses of (c) as-manufactured CNT-PANI electrodes at 0° and at 90° to the principal direction, and (d) in the principal direction after the application of 1000 charge-discharge cycles or pre-charging to $\Phi_{PRE} = 1$ V and $\Phi_{PRE} = -0.2$ V. The (e) Young's modulus and (f) yield strength for CNT mat ($f_P = 0$) and for CNT-PANI composite electrodes as measured in experiment and as predicted with finite element simulation. Finite element predictions are included for PANI volume fractions $f_P = 0$ and $f_P = 0.30$.

# Supplementary Information for

# THE MECHANICAL AND ELECTROCHEMICAL PROPERTIES OF POLYANILINE-COATED CARBON NANOTUBE MAT

Wei Tan[1,2], Joe C. Stallard[1], Changshin Jo[1,3], Michael F. L. De Volder[1], Norman A. Fleck[1]*

[1] Engineering Department, University of Cambridge, Trumpington Street, Cambridge, CB2 1PZ, UK.

[2] School of Engineering and Materials Science, Queen Mary University London, London, E1 4NS, UK.

[3] School of Chemical Engineering & Materials Science, Chung-Ang University, 84 Heukseok-ro, Dongjak-gu, Seoul, 06974, Republic of Korea

*Corresponding author: Email: naf1@eng.cam.ac.uk, Tel: +44 (0) 1223 748240, Fax: +44 (0) 1223 332662.

## S1. Transmission electron microscopy of CNT bundles

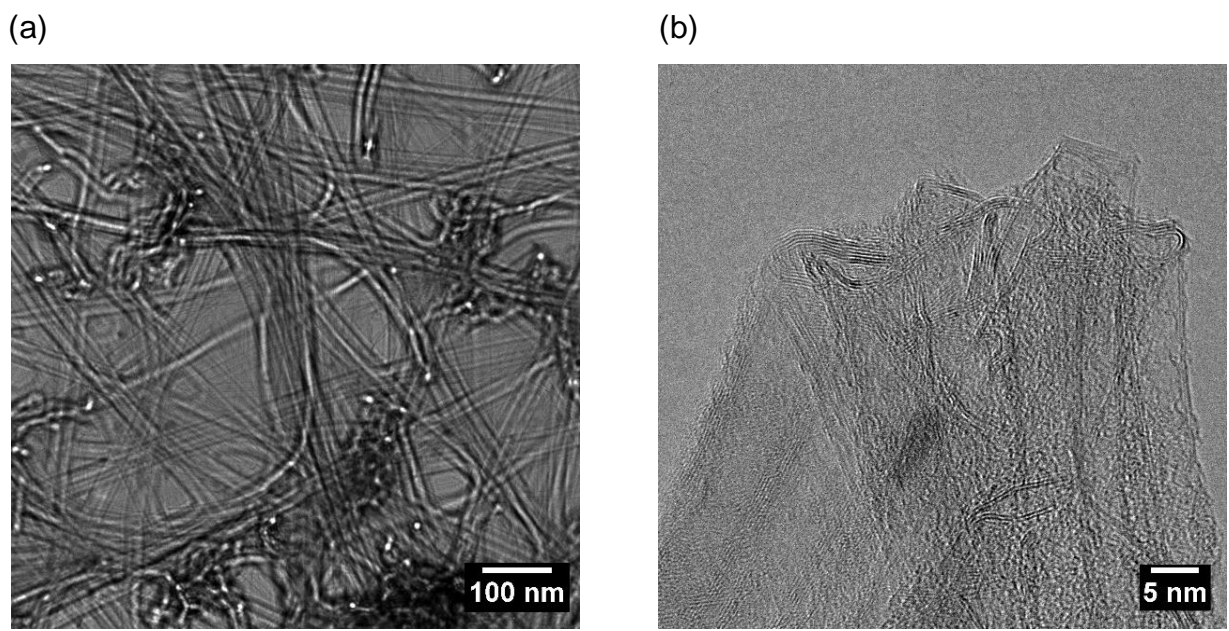

Figure S1: Images of (a) the CNT mat microstructure of interconnected CNT bundles, and (b) of the CNT bundle microstructure, both obtained with transmission electron microscopy.

**S2. Energy-dispersive X-Ray spectra recorded for CNT mat and CNT-PANI-T20 composite electrodes measured after manufacture**

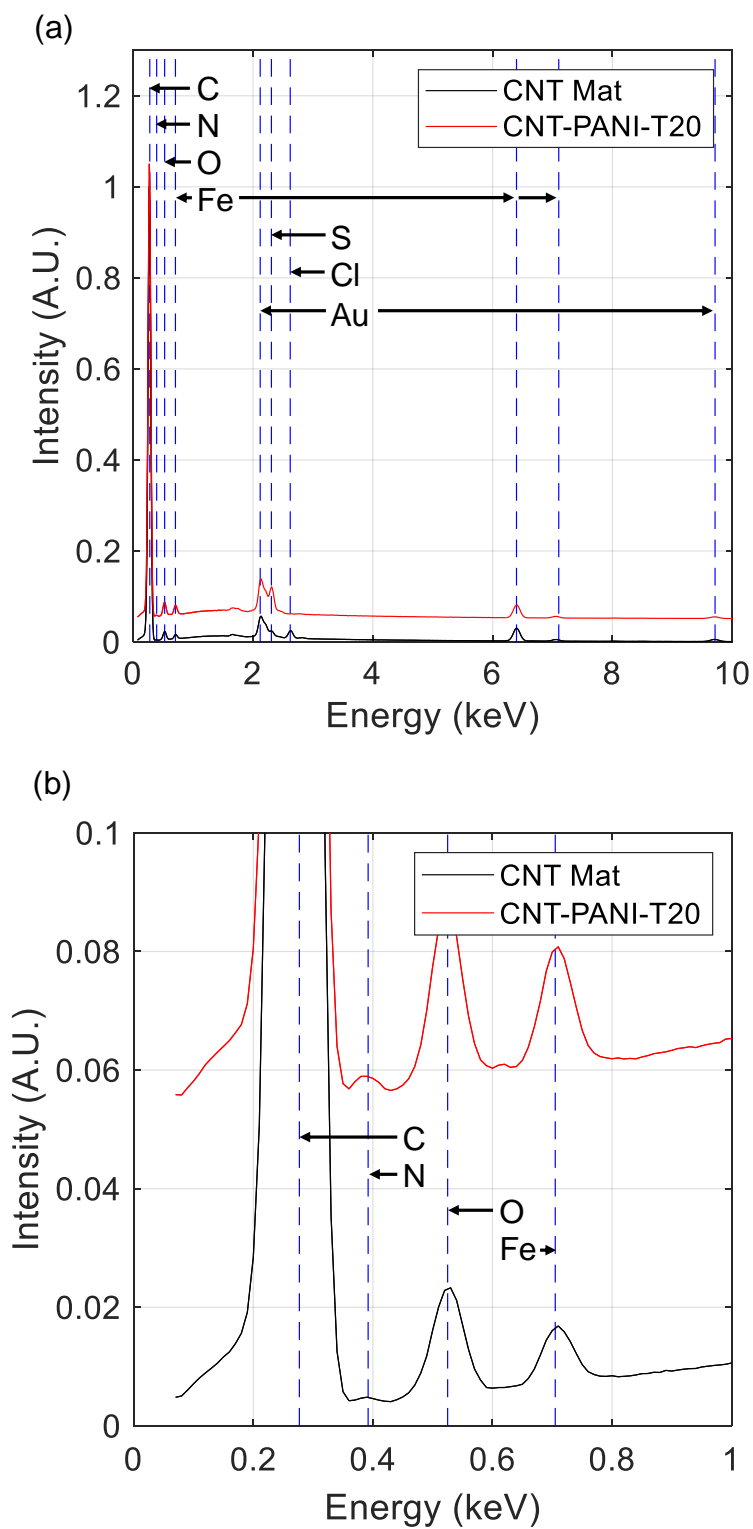

Figure S2: (a) Spectra of CNT mat and CNT-PANI-T20 composite electrode. The plotted intensity is normalised by the highest value within the spectra; that of the CNT-PANI-T20 electrode is shifted upwards by a value of 0.05 units for clarity. The enlarged view shown in (b) reveals a minor peek corresponding to nitrogen in the CNT-PANI-T20 electrode.

**S3. Energy-dispersive X-Ray mapping for CNT-PANI-T20 composite electrodes.**

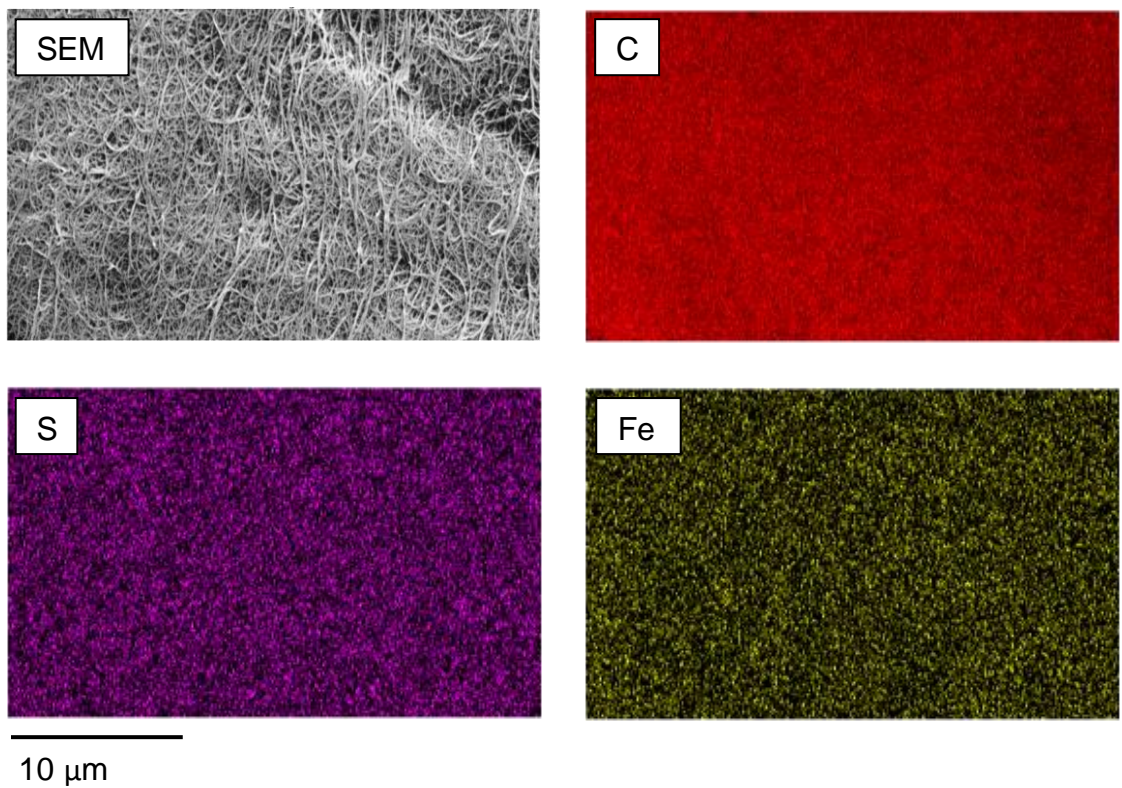

Figure S3: Energy-dispersive X-Ray mapping corresponding to the image of a CNT-PANI-T20 composite electrode microstructure obtained with a scanning electron microscope, for the elements carbon, sulphur and iron.

**S4. X-Ray diffraction spectra of CNT mat and CNT-PANI-T20 composite electrode**

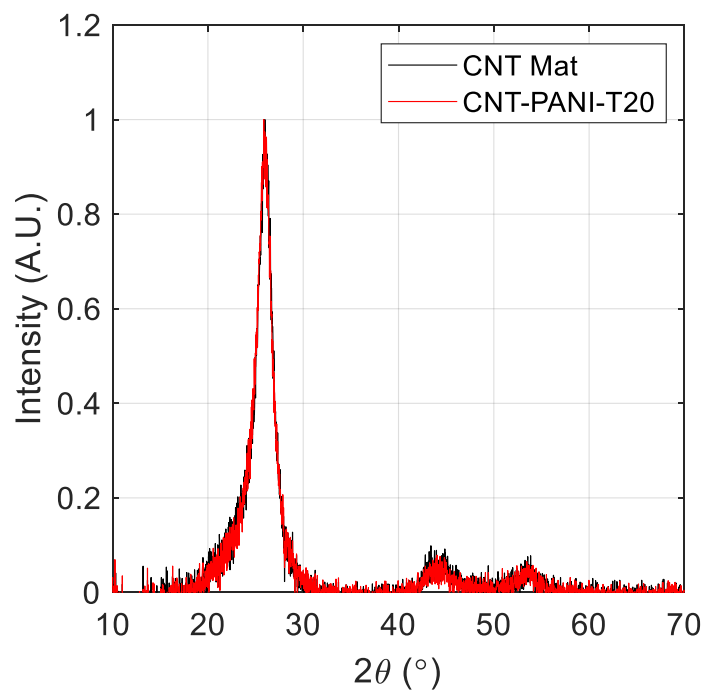

Figure S4: X-Ray diffraction spectra of the CNT mat and CNT-PANI-T20 composite electrode.

## S5. Electrical and mechanical test setup

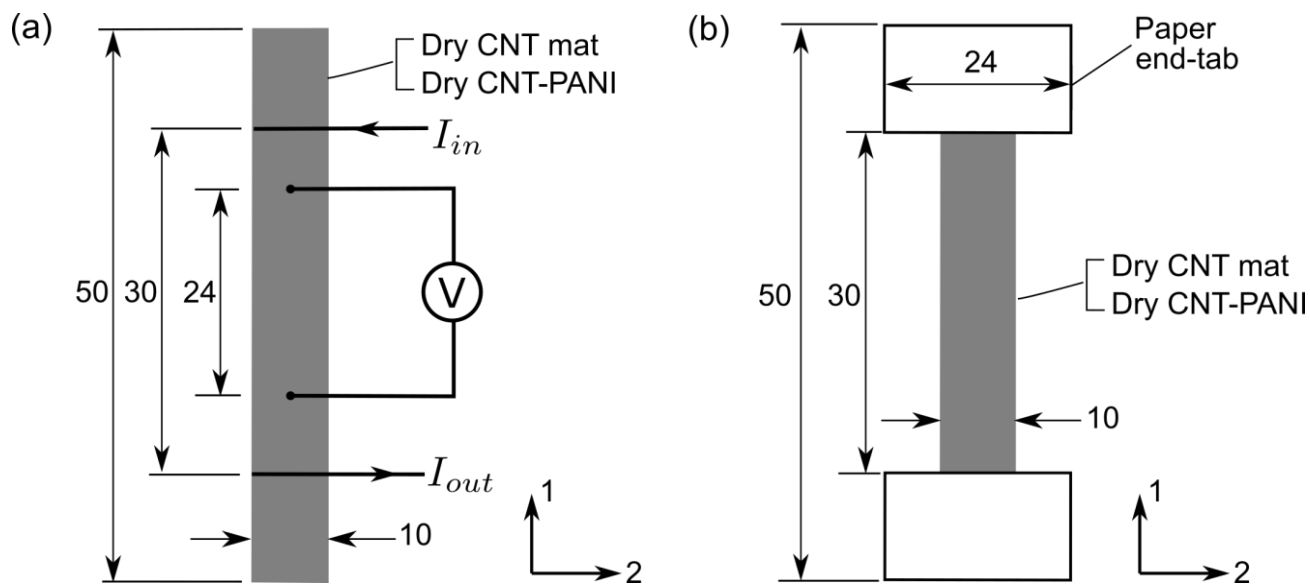

Figure S5: (a) Measurement of electrical conductivity using four-point probe method. (b) Uniaxial tensile test specimen geometries for experiments on CNT mat samples and CNT-PANI composite electrode samples. All dimensions are in mm.

## S6: Effect of cycling upon electrode capacitance

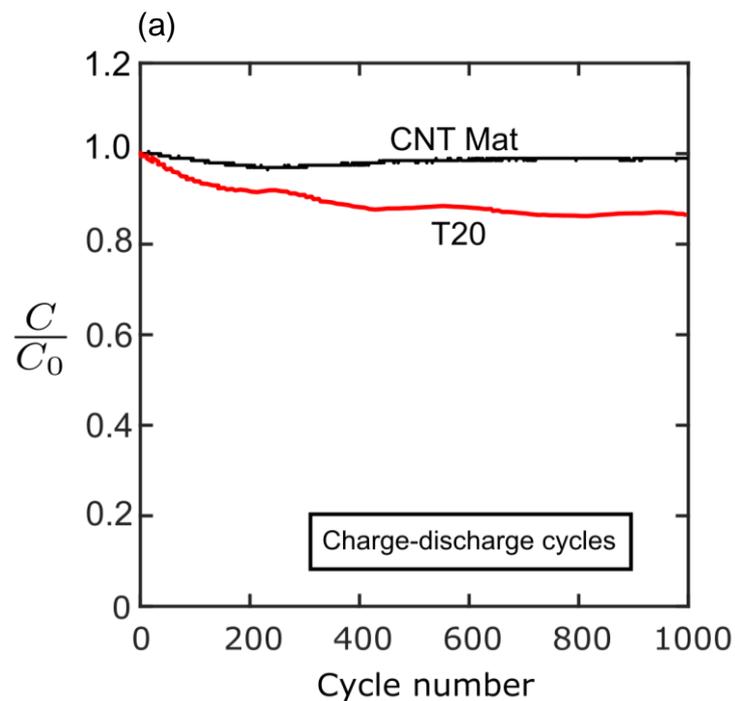

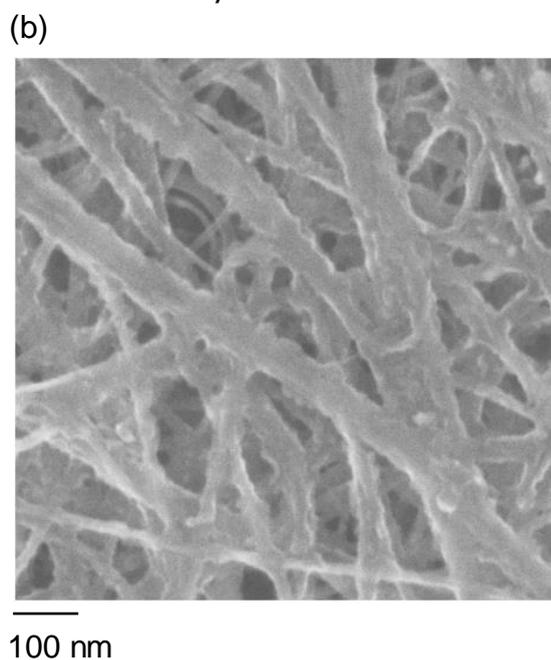

Figure S6: (a) The measured capacitance of a CNT mat and CNT-PANI-T20 electrode plotted against the number of applied charge/discharge cycles. (b) The microstructure of a CNT-PANI-T20 electrode after the application of 1000 charge/discharge cycles, obtained with a scanning electron microscope.

**S7. Raman spectrum**

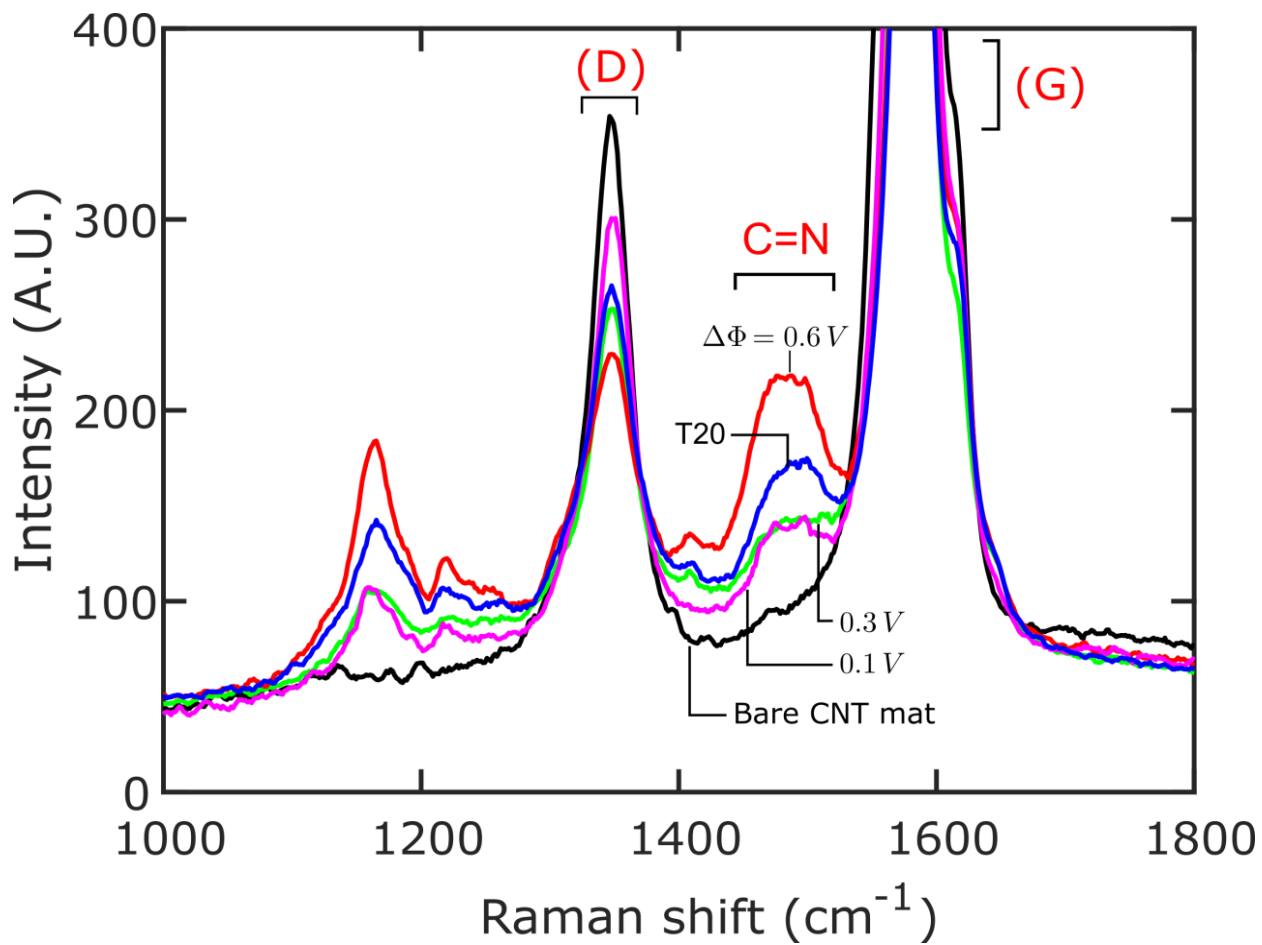

Figure S7: Raman spectra of Bare CNT and CNT-PANI composite electrodes measured after pre-charging to a range of different voltages.

**S8: Boundary conditions and finite element mesh**

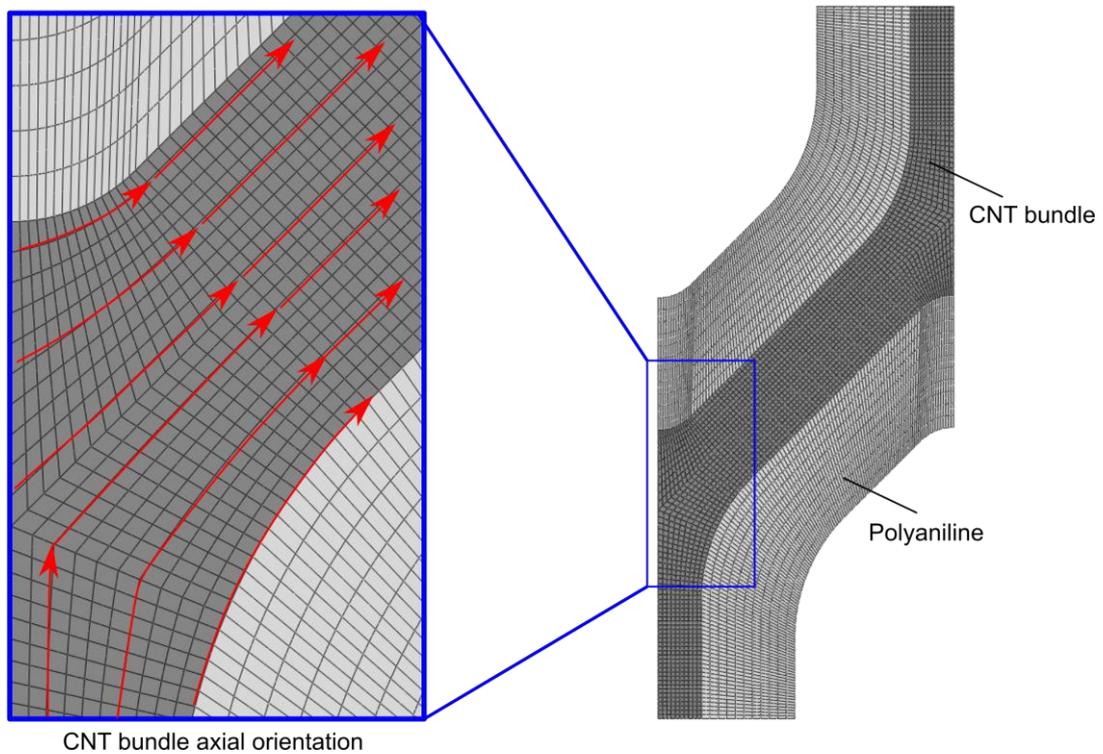

Figure S8: The detail of the finite element mesh used in prediction of the unit cell response; the orientation of the anisotropic CNT bundle material is indicated by the direction of arrows.

## S9: Calibration of the finite element model

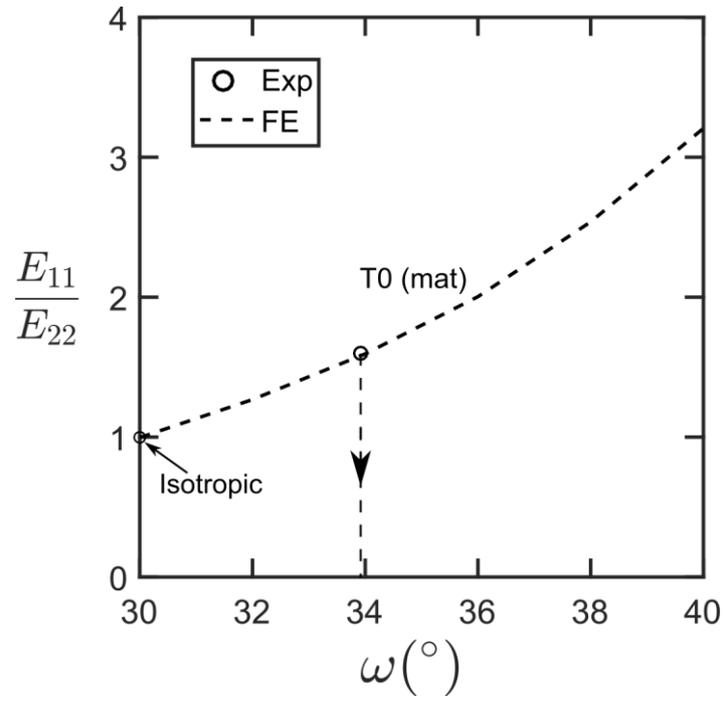

Figure S9: The ratio of the unit cell moduli $E_{11}/E_{22}$ are plotted against the initial angle of inclination of the unit cell $\omega$.